\documentclass[11pt]{JHEP3}
\usepackage{amsmath}
\usepackage{amssymb}
\usepackage{bbold}

\DeclareMathOperator{\Tr}{Tr}
\DeclareMathOperator{\Ad}{Ad}

\title{Symmetry breaking, permutation D-branes on group
manifolds: boundary states and geometric description}

\author{Gor Sarkissian\\
The Abdus Salam Centre ICTP,\\
Strada Costiera 11,\\
34014, Trieste, ITALY\\
\email{gor@ictp.trieste.it}}
\author{Marija Zamaklar\\
MPI/AEI f\"ur Gravitationsphysik\\
Am M\"uhlenberg 1\\
14476 Golm, GERMANY\\ 
\email{marija.zamaklar@aei.mpg.de}}
\keywords{group manifolds, boundary states}
\preprint{AEI-2003-110\\hep-th/0312215}

\abstract{We use the permutation symmetry between the product of
several group manifolds in combination with orbifolds and T-duality 
to construct 
new classes of symmetry breaking branes on products of group manifolds.
The resulting branes  mix the submanifolds and break part of the diagonal chiral
algebra of the theory. We perform a Langrangian analysis as well as a
boundary CFT construction of these branes and find agreement between
the two methods. }

\begin{document}

\section{Introduction and summary}

There are basically two complementary ways of studying D-branes on
curved manifolds:~the microscopic CFT description and the approach
which uses various effective actions (DBI or supergravity). The
relation between these two approaches is non-trivial and it is of
interest to have both descriptions available.  However, many branes
constructed using one or the other method are still lacking the
alternative description.
In some cases, like for classes of D-branes on group manifolds or on
coset spaces, both descriptions of branes are available. The aim of
this paper is to extend this list to a class of new D-branes, on
products of group manifolds. The construction which we present does
not in general require these manifolds to be identical, though most of
our explicit formulas and examples will be given for these cases.

There are two main tools which are used in our construction: the
permutation symmetry between the group manifolds as it was used 
in~\cite{Figueroa-O'Farrill:2000ei,Recknagel:2002qq,Gaberdiel:2002jr,Fuchs:2003yk,Fuchs:1999zi,Fuchs:1999xn,Quella:2002ns}\footnote{In 
the case of a product
of non-identical groups, the permutation symmetry is applied to the
common subgroups of the factor groups.} and the construction of
symmetry breaking branes using orbifolds in combination with T-duality, as 
proposed in~\cite{Fuchs:2000fd,Maldacena:2001ky}.  The resulting branes of our
construction are \emph{symmetry breaking}, i.e.~they preserve less
than the diagonal part of the affine algebra, and are
\emph{non-factorisable} (or  \emph{permutation})  branes, i.e.~they non-trivially mix the factor
groups.

Let us explain the basic idea of our construction on the simplest
example of two identical groups~(i.e.~$H_0=H_1=H$). A generalisation
to several groups will be given in some cases in the main text, where
we will also make the  schematic exposition of the
introduction mathematically precise.  Our starting point is the \emph{maximally symmetric, permutation} (or
\emph{non-factorisable}), brane
of~\cite{Recknagel:2002qq} and~\cite{Figueroa-O'Farrill:2000ei}. Let
us consider the brane whose worldvolume is given by the submanifold of
the~$H \times H$ space defined by
\begin{equation}
\label{b1}
(i): \quad (g_0,g_1)\Big|_{\rm brane}= \Big{\{}(h_{0}f_0h_1^{-1},h_1f_1h_0^{-1})
\,\Big|\,\forall h_i\in H\, ,\,\, (i=0,1)\,  \,  \Big{\}} \, ,
\end{equation} 
where $g_0$ and $g_1$ denote an element of $H_0$ and $H_1$ respectively, $\Big|_{\rm brane}$
denotes the restriction to the brane surface, and $f_0$, $f_1$ are arbitrary, but fixed, elements in~$H$.  One
refers to this brane as a \emph{maximally symmetric brane} since it
preserves the currents
\begin{equation}
\label{b2}
\bar{J}^{a}_0 = J^{a}_1 \, , \quad  J^{a}_0 = \bar{J}^{a}_1 \, \quad
(a=1\ldots  \dim H) \, ,
\end{equation}
i.e.~the full diagonal affine subalgebra~$H_{\text{diag}}^{\bar{(1)},
  (2)} \times H_{\text{diag}}^{(1),\bar{(2)}}$ of the total affine
algebra $H_{(1)} \times \bar{H}_{\bar{(1)}} \times H_{(2)}\times
\bar{H}_{\bar{(2)}}$ .  It is easy to see, using the Sugawara
construction, that the boundary conditions~(\ref{b1}) are conformal.
The reason why  these branes  are called \emph{permutation branes}, is
because the gluing conditions~(\ref{b2}) are obtained from the gluing conditions
for a direct product of branes by permuting the chiral current of one
group with the chiral current of the other group.  This
non-factorisable character of the gluing conditions leads to an
effective geometry of the brane, which turn out to be a submanifold
diagonally embedded in the product of the groups.

The \emph{permutation} symmetry between the subgroups of the target
space is also used in writing the boundary state for these
branes~\cite{Recknagel:2002qq}.  Namely, a schematic form of the
boundary state for this brane is, in the case when the levels of both
groups are the same, given by
\begin{equation}
\label{b3}
|A \mu \rangle_{{\cal P}} = \sum_{\nu,N,M} c_{\mu}{}^{\nu} 
|\nu,N \rangle_0 \otimes\overline{|\nu,N\rangle}_1
\otimes|\nu,M \rangle_1\otimes\overline{|\nu,M \rangle}_0  \,.
\end{equation}
Here $c_{\mu}{}^{\nu}$ are constants and the subscript on the ket
vectors denotes Hilbert spaces in the first and second group
manifolds.  This brane will play a role similar to that of the A-type
brane in the construction of symmetry breaking branes
of~\cite{Maldacena:2001ky}.

Another essential ingredient which we use in this paper is the
construction of the symmetry breaking branes on a group~$H$ using the
relation between the~$Z_k$ orbifold of~$H$ and its T-dual theory.
Namely, it was shown in~\cite{Maldacena:2001ky} that an application of
T-duality on a~$Z_k$ invariant superposition of A-type branes on~$H$
leads to a brane which breaks (a fraction of) the affine algebra and
preserves the currents
\begin{equation}
\label{b4}
J^{Y} = -\bar{J}^{Y} \, , \quad  J^{\alpha} = \bar{J}^{\alpha}
\, \, .
\end{equation}
Here $J^Y$ is the~$U(1)_Y$ current which we T-dualise, $J^\alpha$ are
all the currents in~$H$ which commute with~$J^Y$ and the remaining
currents are not preserved. This is the reason why these branes are
called \emph{symmetry breaking branes}. We will also sometimes call
these branes B-type branes.  It was furthermore shown
in~\cite{Sarkissian:2002ie} that, using the group theory language, the
symmetry breaking branes are described by conjugacy classes multiplied
from the left (or right) by the $U(1)_Y$ subgroup of H, 
\begin{equation}
\label{b45}
(ii): \quad (g)\Big|_{\rm brane}= \Big{\{}Lhfh^{-1}\,
\Big|\,\forall h\in H\, , L = e^{i \alpha Y} \in U(1)_Y \,  \,  \Big{\}} \,   . 
\end{equation}
From (\ref{b4}) we see that the boundary condition in the $U(1)_Y$
direction is Neumann, unlike in the case of the A-type brane from
which this B-brane was derived. Correspondingly, there is an
additional $U(1)$ fibre introduced at each point of a conjugacy class
(i.e.~A-brane) in~(\ref{b45}).

A schematic form of the boundary states for the B-type branes
is~\cite{Maldacena:2001xj}
\begin{equation}
\label{bB}
|B \mu \rangle = \sum_{\nu \in P/kQ^{\vee}}|\mu \nu \rangle \otimes |\nu \rangle'  \, , 
\end{equation}
where $P$ and $Q^{\vee}$ are the weight and coroot lattices of $H$, $k$ is the
level of the group and~$|\mu \rangle'$ is a state in the chiral
algebra~${\cal A}(T)$ of the maximal torus of $H$ which is of B-type
with respect to the~$J^L$ current and of A-type with respect to the
remaining ones.

Starting from the A-type brane (\ref{b1}), we use the logic given
above to write the following two non-factorisable, B-type
permutation branes on a product of group manifolds,
\begin{align}
\label{b5a}
\text{I}&: \; (g_0,g_1)\Big|_{\rm brane}= \Big{\{}(h_0f_0h_1^{-1}, h_1 f_1 h_0^{-1}L_1)
\,\Big|\,\forall h_i\in H\, , (i=0,1)\,  \,  L_1 \in U(1)_{Y_1} \Big{\}} \, , \\[1ex]
\label{b5b}
\text{II}&: \; (g_0,g_1)\Big|_{\rm brane}= \Big{\{}(h_0f_0h_1^{-1}L_0, h_1 f_1 h_0^{-1}L_1)  
\,\Big|\,\forall h_i,\in H\, , \,  \,  L_i \in U(1)_{Y_i}, \, \, (i=0,1) \Big{\}} \, .
\end{align}
These branes can also be considered as product of twisted conjugacy classes
as suggested in \cite{Quella:2002ns}.
By checking the invariance of the WZW action with these boundary
conditions we show that the symmetries preserved by branes~I and~II
are generated by the currents
\begin{alignat}{9}
\label{b6a}
\text{I}&:\quad & \bar{J}_0^{A} &= J_1^{A} \, , &\quad J_{0}^{Y_1} &= -
\bar{J}_{1}^{Y_1}\,  ,  &\quad  J_0^{\alpha} &= \bar{J}_1^{\alpha} \, , \\
\label{b6b}
\text{II}&:\quad & \bar{J}_0^{Y_0} &= - J_1^{Y_0} \, , &\quad  \bar{J}_0^{\alpha} &=
J_1^{\alpha} \, , &\quad J_{0}^{Y_1} &= - \bar{J}_1^{Y_1} \, , &\quad J_0^\alpha &=
\bar{J}_1^\alpha \, . 
\end{alignat}
Here $A$ denotes all generators of $H$ while $\alpha$ and $Y$ are as
stated below formula~(\ref{b4}). We see that the first condition
in~(\ref{b6a}) is the same as for the symmetric, permutational
brane~(\ref{b2}) while the second and the third conditions are
``permuted'' version of the B-type conditions~(\ref{b4}).  For the
second brane, all preserved currents are of the ``permuted'' B-type.
Hence to derive the boundary states for these branes we start from the
boundary state for the maximally symmetric permutation
brane~(\ref{b3}). In order to reduce the symmetry, we apply the
procedure of \cite{Fuchs:2000fd,Maldacena:2001ky}: perform the $Z_k$ orbifold and a T-duality in the second
group (for type~I brane) or in both groups (for type~II branes). The  unbroken currents 
remain related in the permuted way.


The resulting boundary states will have the following schematic form
for the branes~(\ref{b5a}) and~(\ref{b5b}) (in the case of group $H=SU(2)$),
\begin{eqnarray}
\label{Iboun}
|I\rangle &=& \sum_{\tilde{\mu}, N, n} c_{\mu}{}^{\tilde{\mu}}
 |\tilde{\mu} N \rangle_0 \otimes  \overline{|\tilde{\mu} N
 \rangle}_{\bar{1}} \otimes |\tilde{\mu} n\rangle\rangle_{1
 \bar{0}}^{PF} \otimes  {|n \rangle\rangle'}_{1 \bar{0}}^{U(1)} \, \\
\label{IIboun}
|II\rangle &=& \sum_{\tilde{\mu}, n,m} c_{\mu}{}^{\tilde{\mu}} 
|\tilde{\mu} n\rangle\rangle_{0 \bar{1}}^{PF} \otimes  
{|n\rangle\rangle'}_{0 \bar{1}}^{U(1)} 
\otimes |\tilde{\mu} m\rangle\rangle_{1 \bar{0}}^{PF} \otimes  
{|m\rangle\rangle'}_{1 \bar{0}}^{U(1)} \, .
\end{eqnarray}
That is, in the first case, the resulting brane is a product of
permuted A-Ishibashi states with permuted B-Ishibashi states,
while in the second case it is a product of two permuted B-Ishibashi
states.\footnote{For explicit expressions for these boundary states see equations~(\ref{seccar})
and~(\ref{thircar}) in the main text.}  Is is easy to see that the
boundary conditions~(\ref{b5a}) and~(\ref{b5b}) are conformal.  The
total diagonal energy momentum tensor is the same as for the direct
product of~A and~B branes (i.e.~two B-branes). The only effect of the
boundary conditions is to permute (with respect to the case of a
direct product) the way in which the different chiral components of the
energy momentum tensor are related to the anti-chiral components.

Something interesting happens when the $U(1)$ groups in~(\ref{b5b})
are embedded in such a way that $L_0=L_1^{-1}$.
In this case the conditions on the currents with
indices~$\alpha$ are as for the maximally symmetric permutation
brane~(\ref{b2}), the conditions on the currents~$L$ are
``unpermuted'' as
\begin{equation}
\label{b7}
\text{III}: \quad \bar{J}_0^{Y} =  J_0^{Y} \, , \quad  J_1^Y =  \bar{J}_1^Y \, ,
\end{equation}
while the remainder of the currents (with generators which do not
commute with~$Y$) get broken.
The boundary state for this  brane can be deduced from the boundary state for the  maximally symmetric permutation 
brane~(\ref{b3}). One first decomposes the permuted Ishibashi
states~$0 \bar{1}$ and~$1 \bar{0}$ as a product of parafermions
and~$U(1)$ factors, and then ``unpermutes'' the~$U(1)$ Ishibashi
states. The resulting boundary state is
\begin{equation} 
\label{IIIboun}
| {\rm III} \, \mu \rangle = \sum_{\tilde{\mu}, n} c_{\mu}{}^{\tilde{\mu}} |\tilde{\mu} n \rangle\rangle_{0 \bar{1}}^{PF} \otimes
|\tilde{\mu} n \rangle\rangle_{1 \bar{0}}^{PF} \otimes | n \rangle\rangle_{0 \bar{0}}^{U(1)} \otimes  | n \rangle\rangle_{1 \bar{1}}^{U(1)} \, ,
\end{equation}
i.e.~it is a product of permuted parafermionic A-type Ishibashi states
with unpermuted~$U(1)$ A-type Ishibashi states.  

All branes we have considered so far share one common feature: they
are all built from bulk CFT states labeled by the same primaries in
the first and second group, or equivalently, they are all
characterised by one independent label~($f = f_0 f_1$) in the
effective description, as we will show. At the level of the boundary
state this was reflected through the presence of the permuted
parafermionic Ishibashi states, for all boundary states which we
listed.  The opposite situation occurs if one starts with a
\emph{direct} product of conjugacy classes and multiplies them with a diagonally embedded
U(1) group~\cite{Sarkissian:2002nq,Quella:2002ns,Quella:2002fk}
\begin{equation}
\label{b8}
\text{IV}: \quad  (g_0,g_1)\Big|_{\rm brane}= \Big{\{}(L h_0 f_0 h_0^{-1},  L h_1 f_1 h_1^{-1})
\,\Big|\,\forall h_i\in H\, , (i=0,1)\,  \,  L \in U(1)_L \Big{\}} \, .
\end{equation}
In this case the relations between the currents are given by
\begin{equation}
\label{b9}
\text{IV}: \quad J_{0}^Y = -\bar{J}_1^Y\, , \quad \bar{J}_{0}^Y = -J_{1}^Y \, , \quad J_0^{\alpha} = \bar{J}_0^{\alpha} \, , \quad J_1^{\alpha} = \bar{J}_1^{\alpha} \, .  
\end{equation} 
We see that the effect of multiplying with the diagonally embedded
$U(1)$ group is to mix the corresponding $U(1)$ currents in a diagonal
way, while keeping the other currents unpermuted.  Since in this case
mixing between the submanifolds occurs only through the $U(1)$
subgroups, one cannot redefine the group elements so that the brane is
characterised by one label. Hence, in contrast to the previous set of branes,
this brane is characterised by two \emph{independent} labels~$f_0$
and $f_1$.

In order to check the correctness of the boundary states presented
here, we check all Cardy conditions, and 
derive the effective geometries of the branes using the
boundary states.  These are then compared with the corresponding
effective geometries derived from the group-theoretical definitions of
branes.  As required we find agreement between the two approaches.  We
also briefly discuss the cases of non-identical groups in
section~\ref{nonind}.  In these cases, any brane that is not a direct
product of branes on subgroups automatically preserves less than the
maximal diagonal affine subalgebra.

This paper is organised as follows. In sections 2--5 we present
various types of permutation symmetry breaking branes on product of
group manifolds. We analyse their properties, derive the expressions
for the worldvolume fluxes and for some examples we also present
explicit geometries. In the second part of the paper (sections 6 and
7) we construct boundary states for some examples considered in
section 4, and reconstruct the effective geometries of the branes from
the boundary states. We end with comments in section 8.  Finally
we attach four appendices, two of which contain technical details for
some of the calculations and two of which collect useful formulas.

\section{General construction of maximally symmetric, permutation branes}

In this section we review the construction of the maximally symmetric,
permutation branes (\ref{b1}) of~\cite{Figueroa-O'Farrill:2000ei}
and derive some of their properties. A generalisation of this
construction to \emph{non-identical} groups leads to a total breaking
of some of the diagonal affine symmetries, and will be discussed in
section~\ref{nonind}.

\subsection{Definition of the brane}

Let us consider a group manifold $M$, which is a product of $K+1$
copies of a group $G$: $M=G\times \cdots \times G$
\cite{Figueroa-O'Farrill:2000ei}.
  We define the \emph{maximally
symmetric, permutational} brane by the following formula
\begin{eqnarray}
\label{bran}
(g_0,g_1,\cdots, g_K)|_{\rm brane}&=&
\bigg{\{}(h_{0}f_0h_1^{-1},h_1f_1h_2^{-1}, h_2f_2h_3^{-1}\cdots
h_{K-1}f_{K-1}h_K^{-1}, h_{K}f_{K}h_{K+1}^{-1}) \nonumber \\ &\Big|& \, \,
h_0 = h_{K+1} \, , \forall h_i\in G\, , \quad (i=1,\cdots,K+1)\,
\bigg{\}} \, .
\end{eqnarray} 
where $g_i$ denotes an element of the $i$'th copy of $G$ in
target space, and $|_{\rm brane}$ denotes the restriction
to the brane surface.
It is easy to see that by redefinition of the elements $h_i$ one can
always bring an arbitrary brane $(f_0, \dots ,f_K)$ into the form
$(f_{0}f_1 \cdots f_K, e,\dots e)$, where $e$ is the identity
element.\footnote{Note that there is a freedom of rigid (``zero
mode'') motion of the brane on the target space. For example, in the case
of two groups, the equation (\ref{bran}) can be generalised to
\begin{equation}
\label{brane1}
(g_0,g_1)\big|_{\rm brane} = \bigg{\{} \bigg{(} h_0 f_0 h_1^{-1}, (x h_1
x^{-1}) f_1 (y^{-1} h_0^{-1} y)\bigg{)} | \forall h_0, h_1 \in G \bigg{\}} \, .
\end{equation}
Here $x$ and $y$ are two arbitrary but \emph{fixed} elements in $G$,
reflecting a freedom in which one can relate elements in the first and
second group. Different choices for $x$,$y$ lead to configurations
that are related by the ``translations'' on a group $G$,
i.e. different choices of coordinate origin. In order to simplify the
notation we choose to work in a frame $x=y=e$.} Hence, another, more
convenient form of writing the equation (\ref{bran}) is
\begin{multline}
\label{brann}
(g_0,g_1,\cdots, g_K)|_{\rm brane} =\\[1ex]
 \bigg{\{} \big(h_0 f h_0^{-1}
g_K^{-1} \cdots g_{1}^{-1}, g_1, \cdots , g_K\big) \Big| \, \, f \equiv f_0
f_1 \cdots f_k \, , \, \forall h_0, g_i\in G\, , \, (i=1,\cdots,K)
\bigg{\}} .
\end{multline}
The dimension of this brane can easily be determined by looking at the
image of~(\ref{bran}) under the map~$m: M= G \times \dots \times G
\rightarrow G$, defined by $m(g_0,g_1, \dots , g_K)=g_0 g_1\dots g_K
\equiv g$ \cite{Figueroa-O'Farrill:2000ei}. The map $m$ maps the
brane~(\ref{bran}) to the conjugacy class
\begin{equation}
\label{produ}
m(g_0, g_1, \dots , g_k)|_{\rm brane} = {\cal C} = \big\{ C=h_0fh_0^{-1}| h_0 \in G \big\} \, .
\end{equation}
Next, note that the inverse of a point under~$m$ is diffeomorphic
to~$G^{K}$. To see this, observe that for any element of the form
\begin{equation}
h \equiv (f_0 h_1^{-1}, h_1 f_1 h_2^{-1}, h_2 f_2 h_3^{-1}, \dots , h_K^{-1} f_K) \, , \quad  (\forall h_i \in G)
\end{equation}
the relation $m(h) = m(f_0 \dots f_K)$ holds.  Hence altogether we see
that the dimension~$D$ of a generic brane $(f_0,\ldots f_K)$ is given
by
\begin{equation}
\label{size}
D = \dim {\cal C} + K \,\dim G \, .
\end{equation}

Equation~(\ref{bran}) does not fully specify the consistent brane
embedding. Namely, in order to claim that the geometric
embedding~(\ref{bran}) solves the DBI action, one needs to turn on a
gauge invariant worldvolume flux.  Due to the complexity of the DBI
action, we will not determine this flux directly from the equations of
motion. Instead, let us recall that in order to have a well defined
Lagrangian action of the WZW theory on a world-sheet with boundary,
the restriction of the WZW three-form to the D-brane worldvolume has
to belong to the trivial cohomology
class~\cite{Klimcik:1997hp}.\footnote{Generically, there are also
additional global topological restrictions following from the
requirement of independence of the action~(\ref{actwzw}) of the actual
position of the embedding of the auxiliary disk in the group
manifold. These lead, for example, to the quantisation condition for
the position of the~$S^2$ brane in~$SU(2)$. We will not discuss
these kind of conditions here.  The details can be found, for example,
in~\cite{Gawedzki:1999bq}.}
More precisely, there should exist a globally well-defined
two-form~$\omega^{(2)}$ on the brane worldvolume, satisfying the
equation
\begin{equation}
\label{triv}
\omega^{\rm WZ}(g)\big|_{\rm brane} ={\rm d}\omega^{(2)} \, .
\end{equation}
Given the two-form $\omega^{(2)}$, the worldsheet action can be
written as
\begin{eqnarray}
\label{actwzw}
S&=&S(g,k)-{k\over 4\pi}\int_D \omega^{(2)}\, , \\
\label{wzact}
S(g,k)& =& { k\over{4 \pi}}\int_{\Sigma} d^2z 
{\rm Tr}(\partial_zg\partial_{\bar{z}}g^{-1}) + { k\over{4 \pi}}\int_B {1\over 3}{\rm Tr}(g^{-1}dg)^3 \,  \nonumber \\ 
&\equiv& { k\over{4 \pi}} \left[ \int_{\Sigma}d^2z L^{\rm kin} 
+ \int_B \omega^{\rm WZ}\right] \, .
\end{eqnarray}
Here $D$ is an auxiliary disc (mapped under the map~$g$ to the D-brane
submanifold) and joined to the string worldsheet~$\Sigma$ along the
boundary, completing it to the closed manifold. The manifold~$B$ is a
three-manifold satisfying the condition~$\partial B=\Sigma +D$.  The
two-form~$\omega^{(2)}$ is precisely the antisymmetric part of the
matrix appearing in the DBI action~\cite{Bordalo:2001ec},
\cite{Stanciu:2000fz},
\begin{equation}
\label{actdb}
S_{\rm DBI}=\int\sqrt{{\rm det}(G+\omega^{(2)})}\, , \quad \omega^{(2)}={\cal F}=B+F \, .
\end{equation}

Therefore, in order to determine~$\omega^{(2)}$ let us
restrict~$\omega^{WZ}$ to the D-brane surface~(\ref{brann}). Using the
following (Polyakov-Wiegmann) identities
\begin{eqnarray}
\label{pwk}
L^{\rm kin}(g h) &=&  L^{\rm kin}(g) + L^{\rm kin} (h)
 -  \Big({\rm Tr} \big(g^{-1}\partial_z g \partial_{\bar z} h h^{-1}\big)+
 {\rm Tr} \big(g^{-1} \partial_{\bar z} g\partial_z  h h^{-1}\big)\Big)\, , \\[1ex]
\label{pwwz}
\omega^{\rm WZ}(g h) &=& \omega^{\rm WZ}(g) + \omega^{\rm WZ}(h)
 - {\rm d}\Big({\rm Tr} \big(g^{-1} {\rm d}g  {\rm d}h h^{-1}\big)\Big)\, ,
\end{eqnarray}
and the relation~(\ref{produ}) it is easy to see that
\begin{equation}
\label{pwnonfa}
\sum_{i=0}^K \omega^{\rm WZ}(g_i)\Big|_{\text{brane}} =  \omega^{WZ}(C) 
+ {\rm d} \left({\rm Tr}\sum_{i=0}^{K-1}g_i^{-1}dg_id(g_{i+1}\cdots g_K)(g_{i+1}\cdots g_K)^{-1} \right) \, .
\end{equation}
where we have, as before, $C=g_0\dots g_K$.
In deriving this expression one uses 
the identity \mbox{$\omega^{WZ}(g_i^{-1})= - \omega^{WZ}(g_i)$}, which
is crucial for the cancellation of the Wess-Zumino terms involving the
elements~$g_i$.  The first term in equation~(\ref{pwnonfa}) can be
rewritten as a total derivative in a parametrisation independent form,
using the results of~\cite{Alekseev:1998mc}
\begin{equation}
\label{pt2}
\omega^{\rm WZW}(C)={\rm d} \left( {\rm Tr}\big(C^{-1}{\rm d}C{1\over
  1-\Ad_C}C^{-1}{\rm d}C\big) \right) \, .
\end{equation}
Here the operator $(1-\Ad_g)$ (where $\Ad_g$ denotes the adjoint action
on~$G$) is invertible when restricted to the vectors tangent
to~(\ref{produ}).
Putting all ingredients together, we see that Wess-Zumino three-form
reduces to a total derivative on the boundary, as advertised. The
expression~(\ref{pwnonfa}), together with~(\ref{pt2}), gives us an
expression for a covariant worldvolume flux~$\omega^{(2)}$ on the
worldvolume of the D-brane~(\ref{bran}). Since this result is
expressed only in terms of group elements, it is manifestly
\emph{reparametrisation invariant}. Using the
parametrisation~(\ref{bran}), this gauge-invariant form can be written
as
\begin{eqnarray}
\label{explicit}
\omega^{(2)} =  \sum_{i=0}^{K} \Tr\Big(f_i^{-1}h_i^{-1} {\rm
  d}h_if_ih_{i+1}^{-1} {\rm d}h_{i+1}\Big) \, ,
\end{eqnarray}
which 
will be used in the following sections.  Note also that
using~(\ref{pwnonfa}) and~(\ref{pt2}) one can determine~$\omega^{(2)}$
only up to an exact two form, while the boundary equations of motion
fully fix its form.
However, even without solving these equations, one can show
that~$\omega^{(2)}$ given in~(\ref{pwnonfa}) and~(\ref{pt2}) is the
only choice compatible with the geometrical symmetries of the brane,
explored in the following section.

\subsection{Symmetries of the brane}

Next we want to determine the symmetries preserved by the
brane~(\ref{bran}). The boundary conditions~(\ref{bran}) are invariant
under any transformation of the form
\begin{eqnarray}
\label{syme}
\begin{aligned}
g_i &\rightarrow g_ik_i^{-1}\,, &  g_{i+1}&\rightarrow k_ig_{i+1}\,,\\
g_0 &\rightarrow kg_0\,,        &  g_{K}  &\rightarrow g_{K}k^{-1} \,,
\end{aligned}
\quad
 (k_i, k \in G) \,,\quad (i=0,1,\ldots,(K-1))\,,
\end{eqnarray}
which in our parametrisation correspond to the transformations
\begin{equation}
h_{i+1} \rightarrow k_ih_{i+1} \, , \quad h_{k+1} \rightarrow kh_{k+1}\,,
\end{equation}
respectively.  We will now show that the full action
\begin{equation}
\label{actprod}
S=\sum_{i=0}^{k}S(g_i)-{k\over 4\pi}\int_D \omega^{(2)}
\end{equation}
with boundary condition~(\ref{bran}) and~$\omega^{(2)}$ given
in~(\ref{pwnonfa}) is invariant under the following transformations
\begin{align}
\label{sym}
    g_i(z,\bar{z},r) &\rightarrow  g_i(z,\bar{z},r) k_{iR}^{-1}(\bar {z},r)\,,\nonumber\\ 
g_{i+1}(z,\bar{z},r) &\rightarrow  k_{iL}(z,r)g_{i+1}(z,\bar{z},r)\,,  
\quad (k_i\in G) \,, \quad  (i=0,1,\ldots,K-1) \,,\\
k_{iL}(z)\big|_{\rm boundary} &= k_{iR}(\bar {z})\big|_{\rm boundary}=k_i(\tau) 
\,, \nonumber
\intertext{as well as}
\label{symu}
g_0(z,\bar{z},r) &\rightarrow  k_L(z) g_0(z,\bar{z},r)\, ,\nonumber\\
g_k(z,\bar{z},r) &\rightarrow  g_k(z,\bar{z},r)k_R^{-1}(\bar{z},r)\,, \\
k_L(z)\big|_{\rm boundary}&= k_R(\bar {z})\big|_{\rm boundary}=k(\tau) \,, \quad 
(k\in G) \, .\nonumber
\end{align}
Here $z$ and $\bar{z}$ are complex coordinates on the
boundary~$\partial B$ and the coordinate~$r$ is parameterising the
radial direction in the three-ball~$B$. For fixed~$i$, in order to
determine the variation of the action, we only need to consider the
following terms
\begin{eqnarray}
\label{relac}
S(g_i,g_{i+1}) &=& S(g_i)+S(g_{i+1}) \, ,  \\[1ex]
\label{relom}
\omega^{(2)}(h_{i+1}) &=&
\Tr\Big(f_i^{-1}h_{i}^{-1}{\rm d}h_{i}f_ih_{i+1}^{-1}{\rm d}h_{i+1}+
f_{i+1}^{-1}h_{i+1}^{-1}{\rm d}h_{i+1}f_{i+1}h_{i+2}^{-1}{\rm d}h_{i+2}\Big)\, .
\end{eqnarray}
The variation of the kinetic and Wess-Zumino terms in the action can
be read off from~(\ref{pwk}) and~(\ref{pwwz}). Using the fact that,
due to the (anti-)holomorphicity, $\omega^{WZ}(k_{iR/L})=0$, one
deduces that the variation of the Wess-Zumino term reduces to a
surface integral over the disc~$D$ and the string
worldsheet~$\Sigma$. The integral over the string world sheet is
canceled by the corresponding~$\Sigma$ integral coming from the
variation of the kinetic term. The remaining integral over the disc is
\begin{equation}
\label{delreac}
\Delta(S(g_i,g_{i+1}))= -{k\over 4\pi}\int_D\Tr\Big(k_i^{-1}{\rm
  d}k_i\big(g_i^{-1}{\rm d}g_i+{\rm d}g_{i+1}g_{i+1}^{-1}\big)\Big) \, .
\end{equation}
Substituting $g_i=h_if_ih_{i+1}^{-1}$ and $g_{i+1}=h_{i+1}f_{i+1}
h_{i+2}^{-1}$ we obtain
\begin{multline}
\label{chanre}
\Delta(S(g_i,g_{i+1}))=\\ {k\over 4\pi}\int_D\Tr
\Big(k_i^{-1}{\rm d}k_i(h_{i+1}f_{i+1}h_{i+2}^{-1}{\rm d}h_{i+2}
f_{i+1}^{-1}h_{i+1}^{-1}-
h_{i+1}f_{i}^{-1}h_{i}^{-1}{\rm d}h_{i}f_{i}h_{i+1}^{-1})\Big)\,.
\end{multline}
This term is canceled by the variation of the two-form term
in~(\ref{actwzw}).  Computing the change of~(\ref{relom}) we find
\begin{multline}
\label{delrelom}
\omega^{(2)}(k_ih_{i+1})-\omega^{(2)}(h_{i+1}) =\\\Tr
\Big(k_i^{-1}{\rm d}k_i(h_{i+1}f_{i+1}h_{i+2}^{-1}{\rm d}h_{i+2}
f_{i+1}^{-1}h_{i+1}^{-1}-
h_{i+1}f_{i}^{-1}h_{i}^{-1}{\rm d}h_{i}f_{i}h_{i+1}^{-1})\Big)
\end{multline}
which cancels (\ref{chanre}).  The proof of the invariance of the
action (\ref{actprod}) under the variation (\ref{symu}) is similar.

Having determined the symmetries of the brane (\ref{bran}) we can now
turn to the question of which bulk currents are preserved by this
brane. The invariance of the manifold~$M$ under separate left/right
group multiplication in each subgroup gets lifted, on the world sheet
of a \emph{closed string}, to a \emph{local} infinite-dimensional
symmetry group $M(z)\times M(\bar{z})$.  The presence of these
symmetries implies the existence of the conserved currents $J_{i}(z) =
- \partial g_i g_i^{-1}$ and $\bar{J}_i(\bar{z}) = g_i^{-1}
\bar{\partial} g_i$ $(i=0,1,\dots, K)$.  As we have seen, the
symmetries under separate left/right group multiplication are, in the
presence of the worldsheet boundary, reduced to symmetries under
\emph{simultaneous} multiplication~(\ref{sym}) and~(\ref{symu}). This
implies the following relations between the currents,
\begin{alignat}{3}
\label{currcon}
\bar{J}_{i}^a &= J_{i+1}^a\,, &\quad& (i=0,\ldots K-1) \,,  \\
\label{currconn}
\quad J_{0}^a &= \bar{J}_{K}^a \,, && \forall T^a \in {\rm Lie(G)} \, . 
\end{alignat}
We see that all left-moving currents for all groups are identified
with the right moving currents of the ``neighboring'' groups, hence
preserving the diagonal subalgebras~$G_{i, \overline{i+1}}$ of the
affine algebras~$G_i \times \bar{G}_{i+1}$. Note however that the
number of preserved currents is equal to the dimension of the Lie
algebra of the target space, in contrast to the case of non-identical
groups that will be discussed in the next section.

\section{The symmetry-breaking, permutation branes}

Starting from the branes constructed in the previous section, we would
now like to discuss various possibilities for deformations of these
branes, in such a way that the resulting branes are physically acceptable.
All deformations which will be considered lead to breaking of some of
the diagonal affine symmetries of~(\ref{bran}).

\subsection{Deformations by restriction to subgroups}
\label{nonind}
One simple way of breaking the symmetries of (\ref{brann}) is to
impose a restriction on this formula, such that all elements~$g_i$
take value in particular subgroups of~$G$. More precisely, let us
choose subgroups $H_{K}\subset H_{K-1} \cdots H_1 \subset H_0 \equiv
G$, and consider~(\ref{brann}) with the restriction that~$g_i \in
H_i$, $(i=0,1,...,K)$. We choose all~$H_i$ to be proper subgroups
of~$G$ in order to simplify the following analysis. No conceptually
new elements appear if some~$H_i=G$, although some details of the
analysis may be different.
Similar type of constructions have previously appeared 
in~\cite{Sarkissian:2002nq,Quella:2002ns,Quella:2002fk}.

Most of the calculations which were shown in the previous section,
like the determination of the expression for $\omega^{(2)}$ or the
dimension of the brane, go though in this case with small and
straightforward modifications. The only more relevant change involves
the question of preserved symmetries.  It is easy to see that the
restrictions which were imposed change the symmetries of the brane
(i.e.~the full action) from~(\ref{syme}) to
\begin{equation}
\label{nsym1}
\begin{aligned}
g_i&\rightarrow g_ik_i^{-1}\,, &  
g_{i+1}&\rightarrow k_ig_{i+1}\,, &&  (k_i\in H_{i+1}) \, , \, (i=1,\ldots,K-1)\\
g_0&\rightarrow kg_0\,, &
g_k&\rightarrow g_{k}k^{-1}\,,&& (k\in H_K) \, .
\end{aligned}
\end{equation}
There could be additional symmetries present, depending on whether
there is a subgroup $F$ in $G$ which commutes with all subgroups
$H_i$. If this is the case, then there is an additional
symmetry
\begin{equation}
\label{nsym2}
g_0\rightarrow k g_0k^{-1} \, , \quad  g_i \rightarrow g_i \, , \quad
\forall k \in F \, .
\end{equation}
In order to write down the relations between the currents which follow
from~(\ref{nsym1}) and~(\ref{nsym2}) we go to a basis of the Lie
algebra of~$G$ adapted to the chain of subgroups~$H_i$. We denote
by~$T^{a_i}$ the set of all generators of the subgroup $H_i$, 
and with $T^{F}$ the subset belonging to $F$.
Invariance of the modified (\ref{bran}) under the transformations
(\ref{nsym1}) implies that the following combinations of the right and
left affine algebras are preserved,
\begin{equation}
\label{curr1}
\begin{aligned}
\bar{J}_{i}^{a_{i+1}}&= J_{i+1}^{a_{i+1}} \,,\quad (i=0,\ldots,K-1) \,, \quad a_i = 1,\ldots,\dim  H_i\\
\bar{J}_k^{a_k} &= J_0^{a_k} \,.
\end{aligned}
\end{equation}
The symmetry~(\ref{nsym2}) furthermore implies the relation
\begin{equation}
\label{curr2}
J^{F} = \bar{J}^{F} \, .
\end{equation}
We see from~(\ref{curr1}) that all left-moving currents for all groups
(except the~$H_0 \equiv G$) are preserved (via appropriate
identification) while the right moving currents $J_i^{\alpha_i}$ for
($\alpha_i= \dim \, H_{i+1} \dots \dim\,H_i$) are completely removed.

\subsection{Deformation by multiplication with the subgroup}
\label{gr1}
Another way to break some of the affine algebra symmetries was
proposed in \cite{Maldacena:2001ky}: the idea was to construct branes
by applying a T-duality transformation on a~$Z_k$ invariant
superposition of A-type (i.e.~symmetry-preserving) D-branes. In the
Lagrangian formulation this procedure was shown to amount to
multiplication of the conjugacy classes (corresponding to the initial
A-branes) by~$U(1)$ group~\cite{Sarkissian:2002ie}. We will now apply
this logic to the maximally symmetric permutation brane~(\ref{bran}) in order to
generate a new type of symmetry-breaking branes. Since the
brane~(\ref{bran}) has a structure more complicated than that  of a conjugacy class,
there will be several inequivalent ways in which we can implement this idea.

To illustrate the basic idea let us first consider the simplest case
of two identical groups:~$M=G\times G$.  We define the boundary
conditions of the type~I brane as
\begin{equation}
\label{genbr}
{\rm I:} \quad (g_0,g_1)\Big|_{\rm brane} = 
\bigg{\{} (h_0f_0h_1^{-1},\, h_1f_1h_0^{-1}L)\, \Big|\, ,  \, \forall L \equiv e^{i\alpha Y} \in U(1)_Y \bigg{\}}
\end{equation}
where $Y$ is an arbitrary (but fixed) generator in the Cartan subalgebra
of~$G$. As before, in order to fully specify the consistent D-brane  
we need to determine the worldvolume two-form~$\omega^{(2)}$.  We can
reduce this calculation to the one which we did for~(\ref{bran}) by
introducing variables  $K_0=h_0f_0h_1^{-1}$ and $K_1=h_1f_1h_0^{-1}$.  We have already
shown that
\begin{equation}
\label{twtr}
\omega^{\rm WZ}(K_0)\Big|_{\text{brane}}+\omega^{\rm WZ}(K_1)\Big|_{\text{brane}}
={\rm d}\omega^{(2)}(h_0,h_1)\Big|_{\text{brane}}
\end{equation}
with $\omega^{(2)}(h_0,h_1)$ given in (\ref{pwnonfa}).  Using
(\ref{pwwz}) and the property that $\omega^{\rm WZ}(L)=0$ for abelian
groups,\footnote{Actually, this condition can be relaxed to the condition
that~$L$ in (\ref{genbr}) is in the maximal torus of~$G$, while the  extension of $L$  to a nonabelian subgroup is less clear.}  we further get that
\begin{equation}
\label{geco}
\omega^{\rm WZ}(K_1L)\Big|_\text{brane} =\omega^{\rm WZ}(K_1)\Big|_\text{brane}
- {\rm Tr}(K_1^{-1}{\rm d}K_1{\rm d}LL^{-1}) \, .
\end{equation}
Combining (\ref{twtr}) and (\ref{geco}) we finally obtain
\begin{equation}
\label{ngom}
\omega^{\rm WZ}(g)\Big|_{\rm
brane}={\rm d}\omega^{(2)}(h_0,h_1,L)={\rm d}\bigg{(}\omega^{(2)}(h_0,h_1) 
- \Tr (K_1^{-1}{\rm d}K_1{\rm d}LL^{-1}) \bigg{)} \, .
\end{equation}
To determine the symmetries preserved by the brane we first look for
the symmetries preserved by  
the boundary (\ref{genbr}):
\begin{enumerate}
\item
$g_0\rightarrow g_0k^{-1}$, $g_1\rightarrow k g_1$
for all $k\in G$; under this transformation $h_1\rightarrow k h_1$ and $K_1\rightarrow k K_1$.
\item
$g_0\rightarrow k g_0$, $g_1\rightarrow g_1k^{-1}$ for all $k\in G$,
$k\notin U(1)_Y$ and $[k,L]=0$. Under this transformation
$h_0\rightarrow kh_0$.  This means that for example, in the case of
$G=SU(N+1)$ we get that $k\in SU(N)$ generated with isospin generators commuting
with $Y$.
\item
$g_0\rightarrow kg_0$, $g_1 \rightarrow g_1$  for all $k\in U(1)_Y$.
Under this transformation $h_0\rightarrow k h_0$ and $L\rightarrow kL$. 
\item
$g_0 \rightarrow g_0$, $g_1\rightarrow g_1 k$ for all $k\in U(1)_Y$.
Under this transformation $L\rightarrow Lk$.
\end{enumerate}
When extending these transformations to transformations of the action 
(as in equations (\ref{sym}) and (\ref{symu}))
one can show that the \emph{full} action~(\ref{actwzw}) is
invariant separately under the transformations~1 and~2.  On the other
hand, only the following \emph{combination} of the
transformations~3 and~4 is a real symmetry of the full action:
\begin{enumerate}
\item[3'.]
$g_0 \rightarrow k g_0$, $g_1\rightarrow g_1 k$ where
$k\in U(1)_Y$. Under this transformation $h_0 \rightarrow k h_0$, $L\rightarrow kLk$.
\end{enumerate}
We give details of all of these calculation in
appendix~\ref{invariance}.  The set of symmetries listed above implies
that the D-brane~(\ref{genbr}) preserves the following set of
currents,
\begin{alignat}{3}
\label{nbconn}
\bar{J_0}^a &= J_1^a\,,  &\quad&  \forall T^a \in {\rm Lie(G)} \\
\label{nbcon}
J_0^a &= \bar{J_1}^a \,, &\quad& \forall T^a \in {\rm Lie(G)} \, \quad \rm{s.t.} \, \quad  [T^a, Y] = 0 \, , \\
\label{nbconn3}
J_0^Y &=  -\bar{J}_1^Y \, .
\end{alignat}
We see that multiplication of the second group with the~$U(1)_L$
subgroup leads to a removal of some of the currents present in the
symmetric brane~(\ref{currconn}) and, as expected, also flips the sign
of the current in the Y-direction.


\subsection{Generalised symmetry-breaking branes}
\label{gen}
The number of preserved affine symmetries can be further reduced by
implementing the procedure from the previous section on both groups
in \mbox{$M= G \times G$}. More precisely, let us consider the brane
\begin{equation}
\label{genbrr}
{\rm II:} \quad  (g_0,g_1)\Big|_{{\rm brane}} = (h_0f_0h_1^{-1}L_0,\quad h_1f_1h_0^{-1}L_1),
\end{equation}
where $L_0$, $L_1$ belong to two \emph{different} $U(1)$ groups in
$M$: $L_0,\in U(1)_{Y_0}$, $L_1,\in U(1)_{Y_1}$, $L_0=e^{i\beta
Y_{0}}$, $L_1=e^{i\alpha Y_{1}}$.  It is easy to show that this brane
preserves the currents (\ref{nbcon}). On the other hand, the
equations~(\ref{nbconn}) and~(\ref{nbconn3}) get modified in an
obvious manner,
\begin{alignat}{3}
\label{nbconnr}
\bar{J_0}^a &= J_1^a \, , &\quad \forall T^a &\in {\rm Lie(G)} \, \quad \rm{s.t.} \, \quad  [T^a, Y_1] = 0 \, , \\
\label{nbconnrn}
J_0^{Y_1} &=  -\bar{J}_1^{Y_1} \, , &\quad \bar{J}_0^{Y_0} &=  - J_{1}^{Y_0} \, . 
\end{alignat}
A special case occurs when both groups in $M$ are multiplied by the
same $U(1)$ group (i.e.~when we take~$Y_0 = Y_1$ and~$\alpha=-\beta$),
\begin{equation}
\label{genbrrd}
{\rm III:} \quad (g_0,g_1)\Big|_{{\rm brane}} = (h_0f_0h_1^{-1}L,\quad h_1f_1h_0^{-1}L^{-1})\,.
\end{equation}
In this case the action is again invariant under the symmetries~1
and~2 from the previous section (with the restriction in~1 to those
$k$ which commute with the $U(1)$ group). These symmetries lead to the
preserved currents~(\ref{nbcon}) and~(\ref{nbconnr}). The symmetry~3',
however, is replaced by
\begin{enumerate}
\item[3.'']
$g_0\rightarrow kg_0k^{-1}$, $g_1 \rightarrow g_1$, $k\in U(1)_Y$.
Under this transformation $h_0\rightarrow kh_0$  and $L\rightarrow k^{-1}L$.
\item[4.'']
$g_0 \rightarrow g_0$, $g_1\rightarrow kg_1k^{-1}$, $k\in U(1)_Y$ .
Under this transformation $h_1\rightarrow kh_1$,  and $L\rightarrow kL$.
\end{enumerate}
The details of the proof of the invariance of the full action under
these transformations are given in appendix~\ref{symm}.  These new
symmetries imply the following additional relations between the
currents,
\begin{align}
\label{nbconn3d}
J_0^Y &= \bar{J}_0^Y \, , \\
\label{nbconnrnd}
\bar{J_1}^Y &= J_1^Y  \, .
\end{align}
We see that the left and right currents in the (same) Y direction in
both groups are separately related in each group. This is different
from the situation with a diagonal identification of the currents
which occurred in the case of non-identical groups. In that case the
left current from one group (in Y direction) was related to the right
current in the other group~(\ref{nbconnrn}). 

\section{Some examples of non-factorisable branes in $AdS_3 \times S^3 \times S^3 \times S^1$ space}

In this section we will analyse the geometry of the branes constructed
in the previous sections for several explicit examples in the \mbox{$AdS_3
\times S^3 \times S^3 \times S^1$} space. These effective geometries of
branes will be matched in the subsequent sections with the geometry
arising from the boundary state construction of branes.
\bigskip
\vskip10pt
\noindent {\bf Maximally symmetric, permutation branes:} 
\vskip5pt 
\noindent Let us first consider an explicit example of the maximally symmetric
permutation (non-factorisable branes)~(\ref{bran}) for the case in which
the group~$G=SU(2)$. In this case the general formula~(\ref{bran})
reduces to
\begin{equation}
\label{persymm}
(g_0,g_1)\Big|_{{\rm brane}} = (h_0f_0h_1^{-1}, \, h_1f_1h_0^{-1}) \, .
\end{equation}
The preserved currents are
\begin{equation}
J_0^a + \bar{J}_1^a = 0 \, , \quad  J_1^a + \bar{J}_0^a = 0 \,\quad  (a=1,2,3) \, .
\end{equation}
The general expression for two form $\omega^{(2)}$ given
in~(\ref{pwnonfa}) reduces to
\begin{equation}
\label{twofor}
\omega^{(2)}=\Tr\Big(h_0^{-1}{\rm d}h_0(f_0 h_1^{-1}{\rm d}h_1f_0^{-1}-
f_1^{-1}h_1^{-1}{\rm d}h_1f_1)\Big) \, .
\end{equation}
In order to write down the geometry of the brane we will use the
coordinates given in~(\ref{global-s3}) and those given
in~(\ref{standard-s3}) for $SU(2)$.  The image of a generic brane
under the multiplication of elements of the first and second
group is the conjugacy class~(\ref{produ}). Therefore all elements on
the brane surface satisfy the condition
\begin{equation}
\label{brane1-generic}
\begin{aligned}
\Tr(g_0 g_1) = \Tr (f_0 f_1 ) 
&= \cos \theta_0 \cos\theta_1 \cos (\tilde{\phi}_0 + \tilde{\phi}_1) -
\sin \theta_0 \sin\theta_1  \cos (\phi_0 - \phi_1) \\[1ex]
&= \cos \psi_0 \cos \psi_1  - \sin \psi_0 \sin \psi_1 \sin \xi_0 \sin
\xi_1 \cos (\eta_0 -\eta_1)  \\[1ex]
&\quad  +\sin \psi_0 \sin \psi_1 \cos \xi_0 \cos \xi_1 = \text{const}.\,
, \quad |\text{const}| \leq 1 \, .
\end{aligned}
\end{equation}
When $f_0f_1 \neq e$, the conjugacy class (\ref{produ}) is
two dimensional and (\ref{brane1-generic}) is the only equation for the brane surface. In other
words, the brane is five dimensional and it is obvious from
equation~(\ref{produ}) that it is topologically equal to $S^2\times
S^3$; at each point of the image of the map~$m$ given in~(\ref{produ})
there is an~$SU(2) \sim S^3$ fibre~\cite{Figueroa-O'Farrill:2000ei}.
However, when $f_0f_1=e$, the embedding
equations~(\ref{brane1-generic}) are replaced by the stronger set of
conditions $g_0 = g_1^{-1}$, which in the
coordinates~(\ref{global-s3}) and~(\ref{standard-s3}) can be written
as
\begin{equation}
\label{shape1}
\tilde{\phi}_1= - \tilde{\phi}_0 \, , \quad \theta_1 =  \theta_0 \, , \quad \phi_1=\phi_0\pm\pi\, .
\end{equation}
Here the sign in the last
term depends on whether $-\pi\leq\phi_0\leq 0$ or $0\leq \phi_0\leq\pi$ respectively. The equations~(\ref{shape1}) tell us
that the brane is a three sphere embedded in the diagonal and
symmetric way between the two~$SU(2)$ groups.  The
two-form~(\ref{twofor}) vanishes in this case, in agreement with the
observation of~\cite{Figueroa-O'Farrill:2000ei} that any Lie subgroup
of the Lie group is totally geodesic submanifold.
\vskip10pt
\noindent {\bf The symmetry-breaking brane of type I:}
\vskip5pt
\noindent Next we want to determine the geometry of the type~I
brane~(\ref{genbr}) on an~$SU(2)\times SU(2)$ manifold,
\begin{equation}
\label{symbrone}
(g_0,g_1)\Big|_{{\rm brane}} = (h_0f_0h_1^{-1}, h_1f_1h_0^{-1}L),
\end{equation}
where we will take $L$ to be of the form $L = e^{i \alpha {\sigma_3
\over 2}}$. In this case, the preserved currents~\mbox{(\ref{nbconn})--(\ref{nbconn3})} reduce to
\begin{equation}
\label{symI}
J_0^3 = - \bar{J}_1^3 \, , \quad \bar{J}_0^a  = J_1^a \, , \quad (a=1,2,3) \, .
\end{equation}
Under the map $m$ of~(\ref{produ}), the type~I brane gets mapped to
the conjugacy class multiplied by the~$U(1)_{\sigma_3}$ group:
\mbox{$\hat{g}\equiv g_0g_1=h_0f_0f_1h_0^{-1}e^{i\alpha{\sigma_3\over 2}} \equiv
{\cal C} L$}. In what follows we will always denote with 
hats those quantities which appear in a product
of group elements from the first and the second group.
The geometry of the image can be determined as follows
\cite{Sarkissian:2002bg}.  Using the fact
that ${\rm Tr}\, {\cal C}= {\rm Tr} f_0 f_1 = {\rm const} = 2\cos \psi_0$
we can write
\begin{equation}
\label{brgeom}
{\rm Tr}\left(\hat{g}e^{-i\alpha{\sigma_3\over 2}}\right)=2\cos \psi_0 \, .
\end{equation}
From here we see that the element $\hat{g}$ belongs to the image of the
brane surface if and only if there is a~$U(1)$ element
$(e^{i\alpha{\sigma_3\over 2}})$ such that the equation~(\ref{brgeom})
is satisfied.
So let us determine for which $\hat{g}$ this equation admits solutions for $\alpha$.
Denoting with $\hat{\theta}$, $\hat{\tilde{\phi}}$ and $\hat{\phi}$ the
coordinates of~$\hat{g}$ in the parametrisation given in~(\ref{global-s3}),
the equation~(\ref{brgeom}) takes the form
\begin{equation}
\label{brgeom2}
\cos\hat{\theta}\cos(\hat{\tilde{\phi}}-{\alpha\over 2})=\cos \psi_0 \, , 
\end{equation}
or equivalently,
\begin{equation}
\label{brgeom2b}
0 \leq\cos^2(\hat{\tilde{\phi}}-{\alpha\over 2})= {\cos^2 \psi_0 \over \cos^2 \hat{\theta}}  \, \leq 1 \, . 
\end{equation}
Hence, equation (\ref{brgeom2b}) can be solved for $\alpha$ only
when $\cos^2\hat{\theta}\geq \cos^2\psi_0$,
or equivalently when
\begin{equation}
\label{brgeom3}
\cos\hat{\tilde{\theta}}\geq \cos2\psi_0 \ , \quad \hat{\tilde{\theta}} = 2 \hat{\theta} \, .
\end{equation}
We see that the image of the brane is a three-dimensional surface
defined by the inequality~(\ref{brgeom3}). To determine the geometry of
the full brane, let us denote the Euler angles for elements in~$g_0$
and~$g_1$ with ``0'' and ``1'' indices.  Then the
$\hat{\tilde{\theta}}$ and $\hat{\tilde{\phi}}$ angles of their
product are given by~\cite{Vilenkin:1968nk}
\begin{align}
\label{thetpr}
\cos \hat{\tilde{\theta}} &= \cos\tilde{\theta}_0\cos\tilde{\theta}_1-\sin\tilde{\theta}_0\sin\tilde{\theta}_1
\cos(\chi_1+\varphi_0) \, ,  \\
\label{phipr}
e^{i\hat{\tilde{\phi}}} &=
{e^{i{\chi_0+\varphi_1\over 2}} \over \cos {\hat{\tilde{\theta}}\over 2} }
\left(  \cos{\tilde{\theta}_0\over 2} \cos{\tilde{\theta}_1\over 2}e^{i {\chi_1+\varphi_0\over 2}}-
\sin{\tilde{\theta}_0\over 2}\sin{\tilde{\theta}_1\over 2} e^{-i {\chi_1+\varphi_0\over 2}} \right) \, .
\end{align}
Substituting the expression for~$\hat{\tilde{\theta}}$ in the equation for
the image of the brane, we see that a generic brane~(\ref{genbr}) is
six dimensional and given by the inequality
\begin{equation}
\label{thetprun}
\cos\hat{\tilde{\theta}}=\cos\tilde{\theta}_0\cos\tilde{\theta}_1-\sin\tilde{\theta}_0\sin\tilde{\theta}_1
\cos(\chi_1+\varphi_0)\geq \cos 2\psi_0 \, .
\end{equation}
As before, the previous discussion was valid in the cases for which
$f_0 f_1 \neq e$. If~$\psi_0=0$, the conjugacy class ${\cal C}$ is
a point and the total brane is four dimensional, given by the relations
\begin{equation}
\tilde{\theta}_0=\tilde{\theta}_1, \quad \chi_1+\varphi_0=\pi \, .
\end{equation}
\vskip10pt
\noindent {\bf The symmetry-breaking brane of type II:}
\vskip5pt
\noindent As explained before, the symmetries preserved by the brane
of type~I can be broken further by multiplying its first term by
a~$U(1)$ subgroup,
\begin{equation}
\label{nnbr}
(g_0,g_1)\Big|_{{\rm brane}}=\big(h_0f_0h_1^{-1}e^{i\beta{\sigma_3\over 2}},h_1f_1h_0^{-1}e^{i\alpha{\sigma_3\over 2}}\big) \, .
\end{equation}
Here we have taken both $U(1)$ groups to be along the same generator,
but we take them to be parametrised by two \emph{independent}
parameters $\alpha$ and $\beta$. The symmetries of brane I, given
in~(\ref{symI}) are now reduced to
\begin{equation}
\label{symII}
J_0^3 = - \bar{J}_1^3 \, , \quad J_1^3 = - \bar{J}_0^3 \, . 
\end{equation}
Note that the following equation holds
\begin{equation} 
\Tr \Big( g_0 \, e^{-i\beta{\sigma_3\over 2}}\, g_1 e^{-i\alpha {\sigma_3\over 2}} \Big) = \Tr \Big( f_0f_1 \Big) \, .
\end{equation}
Using the same arguments which, in the previous case, led to the
inequality~(\ref{thetprun}), one now concludes that
\begin{equation}
\label{thetprunn}
 \cos\tilde{\theta}_0\cos\tilde{\theta}_1-\sin\tilde{\theta}_0\sin\tilde{\theta}_1
\cos(\chi_1+\varphi_0-\beta)\geq \cos 2\psi_0 \, , 
\end{equation}
where $\Tr(f_0f_1) = 2\cos \psi_0$. As before, the elements $g_0$ and
$g_1$ will belong to the brane surface if and only if
this inequality admits a solution for the parameter $\beta$.
This will happen if and only if the maximum of the left hand side
of~(\ref{thetprunn}) is larger than~$\cos 2\psi_0$. It is easy to see
that this maximum is equal to~$\cos (\tilde{\theta}_0 -
\tilde{\theta}_1)$.  Therefore, the generic brane~(\ref{nnbr}) is
six dimensional and given by an inequality
\begin{equation}
\label{thirbr}
\cos(\tilde{\theta}_0-\tilde{\theta}_1)\geq  \cos 2\psi_0 \, .
\end{equation}
When~$\psi_0 =0$ the brane is five dimensional and given by the equation $\tilde{\theta}_0=\tilde{\theta}_1$.
\bigskip
\vskip10pt
\noindent {\bf The symmetry-breaking brane of type III:}
\vskip5pt
\noindent This brane is derived from the previous one by imposing the
restriction that the parameters $\alpha$, and $\beta$ to satisfy the
relation~$\alpha = - \beta$
\begin{equation}
\label{nnbrd}
(g_0,g_1)\Big|_{{\rm brane}}=\big(h_0f_0h_1^{-1}e^{i\alpha{\sigma_3\over 2}},h_1f_1h_0^{-1}e^{-i\alpha{\sigma_3\over 2}}\big) \, .
\end{equation}
The elements $g_0$ and $g_1$ belong to the brane surface if the
following equation admits a solution for the parameter $\alpha$,
\begin{equation}
\label{nnbrddd}
\Tr\Big(g_0e^{-i\alpha{\sigma_3\over 2}}g_1e^{i\alpha{\sigma_3\over 2}}\Big)=2\cos\hat{\psi_0} \, .
\end{equation}
Using the formulae (\ref{thetpr}) and (\ref{phipr}) we can rewrite this equation as
\begin{equation}
\label{baseq1}
\cos{\hat{\Theta}\over 2}
\cos(\gamma/2-\xi/2-\tilde{\phi}_0-\tilde{\phi}_1)=\cos\hat{\psi_0}\,,
\end{equation}
where
\begin{equation}
\label{thhh}
\cos\hat{\Theta}=\cos\tilde{\theta}_0\cos\tilde{\theta}_1-\sin\tilde{\theta}_0\sin\tilde{\theta}_1\cos\gamma\, ,
\end{equation}
and we have introduced new labels $\gamma=\chi_1+\varphi_0-\alpha$ and
$\xi=\hat{\tilde{\phi}}-{\chi_0+\varphi_1\over 2}$. The variables
$\xi$ and $\gamma$ are related to each other by the equation
\begin{equation}
\label{phiprrr}
e^{i{\xi\over 2}}={ 1 \over \cos{\hat{\Theta}\over 2}} \left( \cos{\tilde{\theta}_0\over 2}\cos{\tilde{\theta}_1\over 2}e^{i{\gamma \over 2}}-
\sin{\tilde{\theta}_0\over 2}\sin{\tilde{\theta}_1\over 2}e^{-i{\gamma\over 2}} \right) \, .
\end{equation}
Hence the brane consists of those points for which
equation~(\ref{baseq1}) admits a solution for~$\alpha$.  For~$\psi_0=0$
there are again additional constraints, which imply that in this case
the brane is four dimensional and given by the equations
\begin{equation}
\label{nnbrdd}
\tilde{\theta}_0=\tilde{\theta}_1, \quad \tilde{\phi}_0+\tilde{\phi}_1=\pi \, .
\end{equation}
\vskip10pt
\noindent {\bf Restriction to the subgroups:} 
\par\nopagebreak\vskip5pt 
\noindent As was explained in the previous section, the restriction to
subgroups of~$G$ for different factors in~(\ref{bran}) leads to total
breaking of some of the affine symmetries.  In particular, let us
consider an~$SU(2)\times U(1)$ manifold and a brane
\begin{equation}
\label{suuper}
(g_0,g_1)\Big|_{\rm brane}=\big(h_0 f_0 h_0^{-1} e^{-i\eta {\sigma_3 \over 2}}, e^{i \eta {\sigma_3 \over 2}}\big)\,,
\end{equation}
where $h_0,f_0 \in SU(2)$. 
Following the steps elaborated on around formula~(\ref{thetpr}) one
can deduce that the embedding of this brane in the case in which $f_0
\neq e$ is given by
\begin{equation}
\label{rela}
\cos \theta\cos(\tilde{\phi}-{\eta\over 2})= \cos(\psi_0) \,  ,
\end{equation}
where $2\cos\psi_0=\Tr f_0$, while $\theta$ and $\tilde{\phi}$ are
coordinates on the~$S^3$ in the parametrisation given
in~(\ref{global-s3}). The difference with respect to the previous
cases is that $\eta$ in this case is a coordinate in the $U(1)$
subgroup and not a parameter. Hence the brane surface is defined by
equation~(\ref{rela}) and it is three dimensional.
When $f = e$ the brane is a one-dimensional $U(1)$ brane, diagonally
wrapping a two torus which is the direct product of~$U(1)_{\sigma_3}\times
U(1)$.

Let us also consider a more complicated example of the
restricted~(\ref{bran}) branes on an~$M=SU(2)\times SU(2)\times U(1)$
manifold. The brane is given by
\begin{equation}
\label{brane2}
(g_0,g_1,g_2)\Big|_{\rm brane} = \bigg{\{ } \big( h_0 f h_1^{-1}, \,
h_1 h_0^{-1} e^{-i \eta {\sigma_3 \over 2}}, \, e^{i \eta {\sigma_3 \over 2}} \big) \Big| h_0, h_1, f \in SU(2) , \, g_2 \in U(1) \bigg{\}}\,,
\end{equation}
with the two form $\omega^{(2)}$ given by (\ref{pwnonfa}) and (\ref{pt2}).
The conserved currents are
\begin{equation}
\label{glue2}
\begin{aligned}
J_0^3&=\bar{J}_2\,, \\
\bar{J}_0^a&= J_1^a\quad (a=1,2,3)\,,\\
\bar{J}_1^3&= J_2\,.
\end{aligned}
\end{equation}
To determine the geometry of this brane, we note that the first two
factors in~(\ref{brane2}) define the brane of type~I~(\ref{symbrone}),
which is connected with the $U(1)$ factor in a diagonal way, as
in~(\ref{suuper}). When $f \neq e$, the brane is six dimensional with the
geometry given in Euler coordinates by
\begin{equation}
{\rm Tr} (g_0 g_1 g_2) = \cos\hat{\theta}\cos(\hat{\tilde{\phi}}-{\eta \over 2})=\cos \psi_0 \, ,
\end{equation}
where $\hat{\theta}$ and $\hat{\tilde{\phi}}$ are defined in
equations (\ref{thetpr}) and (\ref{phipr}), respectively.  In the
special case in which $f = e$, the brane equation is $g_0 = (g_1
g_2)^{-1}$ which, when written in components, gives
\begin{equation}
\varphi_0+\chi_1=\pi \, , \quad \tilde{\theta}_0 = \tilde{\theta}_1 \, , \quad  \pi+\chi_0 + \varphi_1 = - \eta \, .
\end{equation}
So we see that in this special case the brane is four dimensional.

\section{Boundary states for maximally symmetric, permutation  branes }
\label{sympres}
In this and the following sections we will present boundary states for
the D-branes considered in the previous sections. We will always work
in the closed string channel, and in cases of several groups we will
choose the levels of all groups to be the same. 
To set up the scene, we start the discussion by
reviewing the construction of the boundary states for maximally symmetric,
permutation branes presented
in~\cite{Recknagel:2002qq}. Using the constructed boundary states we then calculate the
effective geometries of these branes, recovering the classical results from the
previous sections.  In the section \ref{bs}, we then continue by
constructing the boundary states and effective geometries for cases of
symmetry-breaking non-factorisable branes.

\subsection{Construction of the boundary state}
 
To construct the boundary states for maximally symmetric, permutation
branes on $SU(2)_k\times SU(2)_k$ manifold, one starts with
the boundary state for a direct product of two maximally symmetric
branes on~$SU(2)$. These branes preserve the diagonal affine algebras
in both groups \emph{separately}
\begin{equation}
\label{br-pr}
J_0^a+\bar{J}_0^a = 0\, , \quad  J_1^a+\bar{J}_1^a = 0 \, , \quad a=1,2,3 \, .
\end{equation}
The boundary state is described by the tensor product of the
corresponding Cardy states for the first and the second groups
\begin{equation}
\label{carst}
|a_0,a_1\rangle=|a_0\rangle_C^{SU(2)_0}\otimes|a_1\rangle_C^{SU(2)_1}\,, 
\end{equation}
where the boundary states for each group are of the standard form 
\begin{equation}
\label{sucar}
|a_i\rangle_C^{SU(2)_i}=\sum_{j_i}{S_{a_ij_i}\over \sqrt{S_{0j_i}}}|A,j_i\rangle\rangle_u^{SU(2)_i} \, , \quad (i=0,1) \, ,
\end{equation}
and the Ishibashi states are given by
\begin{equation}
\label{ishsu}
|A,j_i\rangle\rangle_u^{SU(2)_i}=\sum_{N}|j_i,N\rangle_i\otimes\overline{|j_i,N\rangle}_i \, , \quad (i=0,1) \, . 
\end{equation}
As usual $S_i{}^j$ is the matrix of a modular transformation $\tau
\rightarrow - {1\over \tau}$ and $|j_i,N\rangle_i$ is 
an orthonormal basis of the irreducible representation $j_i$ of
$SU(2)_k$, $j_i= (0, {1\over
2},1,...{k\over2})$. We will use the following
notation for the boundary states in the rest of the paper:~the
subscript $u$ will indicate that the Ishibashi state is formed from
the left and right states of the same theory, the subscript~$\tau$
will denote that the state is formed from different theories, and a
superscript will indicate which Hilbert spaces are used. All Cardy
states will be denoted with $|\rangle_C$.

As discussed in section 2, one can construct the symmetry preserving,
permutational brane~(\ref{persymm}) using a permutation symmetry (${\cal
P}(h_0,h_1)=(h_1,h_0)$)
 to twist the product of two conjugacy classes.  In this simple case,
the twisting changes the the relation between the currents
(\ref{br-pr}) to
\begin{equation}
\label{preserv}
J_0^a + \bar{J}_1^a = 0 \, , \quad \bar{J}_0^a + J_1^a = 0 \, .
\end{equation}
The preserved currents (\ref{preserv}) imply that the allowed Ishibashi
states for the permutational branes (denoted with the subscript ${\cal
P}$) are of the form
\begin{equation}
\label{permish}
|j_0,j_1\rangle\rangle_{{\cal P}}=|j_0\rangle\rangle_{\tau}^{SU(2)_0\times SU(2)_{\bar{1}}}
\otimes|j_1\rangle\rangle_{\tau}^{SU(2)_1\times SU(2)_{\bar{0}}}
\end{equation}
where
\begin{eqnarray}
\label{permish1}
|j_0\rangle\rangle_{\tau}^{SU(2)_0\times SU(2)_{\bar{1}}}
=\sum_{N}|j_0,N\rangle_0\otimes\overline{|j_0,N\rangle}_1 \, , \\
\label{permish2}
|j_1\rangle\rangle_{\tau}^{SU(2)_1\times SU(2)_{\bar{0}}}
=\sum_{M}|j_1,M\rangle_1\otimes\overline{|j_1,M\rangle}_0 \, .
\end{eqnarray}
Note that not all of Ishibashi states (\ref{permish}) can be used to
construct the boundary state. This is because the boundary state is
part of a closed string Hilbert space, and hence, since the bulk
partition function is diagonal, only Ishibashi states with~$j_0=j_1$
will be allowed.  Notice however that the indices~$M$ and~$N$
in~(\ref{permish1}) and~(\ref{permish2}) are independent.
The complete boundary state is a linear combination of allowed
Ishibashi states. The requirement that Cardy's consistency conditions
hold restrict the allowed linear combination to
\begin{equation}
\label{carsttw}
|a\rangle_{{\cal P}}=\sum_{j}{S_{aj}\over S_{0j}}|j,j\rangle\rangle_{{\cal P}}=\sum_{j}{S_{aj}\over S_{0j}}
\sum_{N,M}|j,N\rangle_0\otimes\overline{|j,N\rangle}_1\otimes|j,M\rangle_1\otimes\overline{|j,M\rangle}_0
\, .
\end{equation}
The proof of the Cardy conditions for this brane is given in
appendix~\ref{cardyproof}.  Note also that although the permutational
brane lives on a product of two manifolds, the boundary state is
characterised by a \emph{single} primary~$a$, in contrast to the
untwisted brane~(\ref{carst}). This fact is in agreement with the
statement that the brane~(\ref{bran}) can always be put into the
form~(\ref{brann}), characterised by a \emph{single} group
element~$g_0g_1$.

\subsection{The effective geometry of the brane}

Given a boundary state, the shape of the brane can be deduced by
considering the overlap of the boundary state with the localised bulk
state $|\vec{\theta} \rangle$, with $\vec{\theta}$ denoting 
the three $SU(2)$ angles in some coordinate system
\cite{Felder:1999ka,Maldacena:2001ky,Schomerus:2002dc}. As we will
see, the boundary state wave function over the configuration space of
all localised bulk states peaks precisely at those states which are
localised at positions derived by the effective methods in the
previous sections.  In the large~$k$ limit, the eigen-position bulk
state is given by
\begin{equation}
\label{ishov}
|\vec{\theta} \rangle = \sum_{j,m,m'} \sqrt{2j+1} {\cal D}_{m m'}^j
 (\vec{\theta}) |j,m,m' \rangle \, ,
\end{equation}
where ${\cal D}_{m m'}^j$ are the  Wigner ${\cal D}$-functions:
\begin{equation}
{\cal D}_{m m'}^j=\langle jm|g(\vec{\theta})|jm'\rangle, \quad \langle jm|jm'\rangle=\delta_{m,m'}
\end{equation}
where $|jm\rangle$ are a basis for the spin $j$ representation of $SU(2)$.
To calculate the overlap with the boundary state, we will need the
knowledge of $S$-matrix of $SU(2)$ at level $k$,
\begin{equation}
\label{smatr}
S_{aj}=\sqrt{{2\over k+2}}\sin\left({(2a+1)(2j+1)\pi\over k+2}\right)
\, .
\end{equation}
In the large-$k$ limit 
the ratio of S-matrix elements appearing in the boundary state simplifies to
\begin{equation}
\label{lark}
{S_{aj}\over S_{0j}}\sim {(k+2)\over \pi (2j+1)}\sin[(2j+1)\hat{\psi}] \, ,
\end{equation}
where, to shorten the notation, we have introduced
 $\hat{\psi}={(2a+1)\pi\over k+2}$.  Using these results, the overlap
 between the boundary state and the localised bulk state becomes
\begin{equation}
\label{twov}
\langle\vec{\theta}_0,\vec{\theta}_1|a\rangle_{{\cal P}}\sim
\sum_{j,m,n}{(k+2)\over \pi}\sin[(2j+1)\hat{\psi}]
{\cal D}^j_{nm}(g_0(\vec{\theta}_0)){\cal
D}^j_{mn}(g_1(\vec{\theta}_1))  \, .
\end{equation}
To simplify this expression we need the identity 
\begin{equation}
\label{matprod}
\sum_m{\cal D}^j_{nm}(g_0(\vec{\theta}_0)){\cal D}^j_{mn'}(g_1(\vec{\theta}_1))
={\cal D}^j_{nn'}(g_0(\vec{\theta}_0)g_1(\vec{\theta}_1)) \, , 
\end{equation}
which follows from the fact that the matrices ${\cal D}^j_{nm}$ form
a representation of the group.
Finally, one needs the property of the Wigner D-functions that
$\sum_n{\cal D}^j_{nn}(g)={\sin(2j+1)\psi\over \sin\psi}$, where
$\psi$ is the angle of the standard metric~(\ref{standard-s3}) and
defined by the relation ${\rm Tr}g = 2 \cos \psi$ (or in our case ${\rm
Tr}(g_0 g_1) = 2 \cos \psi$). The overlap (\ref{twov}) becomes
\begin{equation}
\label{twovtr}
\langle\vec{\theta}_0,\vec{\theta}_1|a\rangle_{{\cal P}} \sim {k+2
\over \pi\sin \psi }\sum_j \sin[(2j+1)\hat{\psi}]\sin[(2j+1)\psi] 
\end{equation}
and from the completeness of $\sin(n\psi)$ on the interval $[0,\pi]$
one concludes
\begin{equation}
\label{fresu}
\langle\vec{\theta}_0,\vec{\theta}_1|a\rangle_{{\cal P}} \sim {k+2
\over 4\sin \psi }\delta(\psi - \hat{\psi})\, .
\end{equation}
Hence we see that the brane wave function is localised on~\mbox{$\psi =
\text{const}.$} bulk states, which is indeed the same relation as the one
obtained in the effective approach~(\ref{brane1-generic}).

\section{Boundary states for symmetry breaking branes}
\label{bs}
Starting from a direct product of maximally symmetric branes and using
a permutation symmetry of the theory, we have in the
previous section generated maximally symmetric permutation branes. In this
section we will use these branes and apply the 
technique of~\cite{Maldacena:2001ky} to generate symmetry breaking,
permutation branes of type~I and type~II. In section~\ref{md} we will first review the
essential steps of the MMS construction, which we will 
then apply in section~\ref{smbs} and \ref{brtwo}.
Finally, in section~\ref{bsiii},
we will use the permutation symmetry between the~$U(1)$ subgroups of
the $SU(2)\times SU(2)$ group in order to derive the boundary state
for brane~III, starting from the direct product of two~$SU(2)$, A-type
branes.

\subsection{Background material}
\label{md}

Let us start by reviewing the T-duality between a Lens space and the
$SU(2)$ theory.  Geometrically, a Lens space is obtained by quotienting
the group manifold by the right action of the subgroup $Z_k$ of the
$U(1)$, and in the Euler coordinates it corresponds to the identification
$\varphi\sim \varphi+{4\pi\over k}$.  In terms of the $SU(2)$ WZW
model this is the orbifold $SU(2)/Z^{R}_k$, where $Z^{R}_k$ is
embedded in the right $U(1)$. The partition function for this theory
is
\begin{equation}
\label{partlens}
Z=\sum_j\chi_j^{SU(2)}(q)\chi_{jn}^{PF}(\bar{q})\psi^{U(1)}_{-n}(\bar{q})
\end{equation}
and coincides with the one for the $SU(2)$ group, up to T-duality.  This relation
enables one to construct new D-branes in the $SU(2)$ theory starting
from the known ones.  As a first step one constructs the brane in the
Lens theory.  As is usual for orbifolds, this is achieved by summing
over images of D-branes under the right $Z_k$
multiplications. Performing then the T-duality on the Lens theory
brings us back to the $SU(2)$ theory and maps the orbifolded brane to
a new $SU(2)$ brane.

As warm-up exercise let us recall how this procedure works in the case
of a \emph{single} $SU(2)$ group~\cite{Maldacena:2001ky}.  Our
starting point is a maximally symmetric A-brane, preserving the
symmetries~(\ref{br-pr}).  If we shift the brane by the right
multiplication with some element $\omega^l=e^{{2\pi l i\over k}\sigma_3} $
of the~$Z_k^R$ group, then the symmetries preserved by this brane are
\begin{equation}
\label{symmetr}
J^a+ \omega^l \bar{J}^a \omega^{-l} =0\, ,\quad (a=1,2,3) \, , 
\end{equation}
while the brane is described by the Cardy state with rotated Ishibashi state
\begin{equation}
\label{twbr}
|A,a\rangle_C^{\omega^l}=\sum_j {S_{aj}\over \sqrt{S_{0j}}}\sum_N |j,N\rangle\otimes (\omega^l \overline{|j,N\rangle})\, .
\end{equation}
Summing over the images one obtains a $Z_k^R$ invariant state, present in the Lens theory 
\begin{equation}
\label{invariant}
\sum_{l=0}^k|A,a\rangle_C^{\omega^l}=\sum_j {S_{aj}\over \sqrt{S_{0j}}}\sum_{l=0}^k
\sum_N |j,N\rangle\otimes (\omega^l \overline{|j,N\rangle}) \, .
\end{equation}
To compute the sum of the Ishibashi states on the right-hand side, one
next uses the orbifold decomposition of $SU(2)_k$
\begin{equation}
\label{decom}
SU(2)_k=({\cal A}^{ PF(k)}\otimes U(1)_k)/Z_k  \, .
\end{equation} 
This decomposition implies that Ishibashi states for the maximally
symmetric A-brane~(\ref{ishsu}) can be written as
\begin{equation}
\label{amax}
|A,j \rangle\rangle^{SU(2)}= \sum_{n =1}^{2k} {1 + (-1)^{2j+n} \over 2}|A, j,  n \rangle\rangle_{u}^{PF}\otimes
 |A,n \rangle\rangle_{u}^{U(1)} \, ,
\end{equation}
where
\begin{equation}
\label{ishfru}
|A, j, n \rangle\rangle_{u}^{PF }=
\sum_{N}|j, n, N\rangle \otimes\overline{|j, n, N\rangle} \, ,  
\end{equation}
and
\begin{equation}
\label{untwishd}
|Ar \rangle\rangle_{u}^{U(1)}= \exp\left[
\sum_{n=1}^{\infty}{\alpha_{-n} \tilde{\alpha}_{-n}\over n}\right]
\sum_{l \in Z}|{r +2k l \over \sqrt{2k}}\rangle \otimes\overline{|{r +2 kl \over \sqrt{2k}}
\rangle}  \, ,  
\end{equation}
are the A-type Ishibashi states for the parafermion and $U(1)_k$ theories.  
If the $Z_k^R$ subgroup lies in the $U(1)$ group appearing in the
decomposition~(\ref{decom}), then under the action of element $\omega^l \in Z_k^R$ the
expression~(\ref{amax}) transform as
\begin{equation}
 |A,j \rangle\rangle^{SU(2)} \rightarrow 
\sum_{n =1}^{2k} {1 + (-1)^{2j+n} \over 2}\omega^{l n}|A, j, n\rangle\rangle_{u}^{PF}\otimes
 |A,n\rangle\rangle_{u}^{U(1)} \, .
\end{equation}
Hence summing over images projects onto the $Z_k^R$-invariant Ishibashi
states for which  $n$ is restricted to the two values~$0$ and~$k$.
Performing T-duality, flips the sign of
the right moving~$U(1)$ sector and one gets a B-type Ishibashi state
of the original~$SU(2)$ theory,
\begin{multline}
\label{bbsymish}
|B,j \rangle\rangle^{SU(2)}=\\[1ex] \left[{1 + (-1)^{2j} \over 2}|A, j, 0\rangle\rangle_{u}^{PF}\otimes
 |B,0\rangle\rangle_{u}^{U(1)}+{1 + (-1)^{2j+k} \over 2}|A, j, k\rangle\rangle_{u}^{PF}\otimes
 |B,k\rangle\rangle_{u}^{U(1)}\right] \, ,
\end{multline}
where 
\begin{equation}
\label{untwishn}
|Br \rangle\rangle_{u}^{U(1)}= \exp\left[
-\sum_{n=1}^{\infty}{\alpha_{-n}\tilde{\alpha}_{-n}\over n}\right]
\sum_{l \in Z}|{r +2kl \over \sqrt{2k}}\rangle \otimes\overline{|-{r +2k l \over \sqrt{2k}}
\rangle} \,  , 
\end{equation}
is a B-type Ishibashi state of $U(1)_k$ theory satisfying the Neumann
boundary conditions.  Knowing the T-dual expression of
the~(\ref{invariant}) allows one to write down the boundary state for
the B-type brane
\begin{equation}
\label{mmsbbr}
|B,a \rangle_C^{SU(2)}=\sum_{j\in Z}{\sqrt{k}S_{a j}\over \sqrt{S_{0j}}}|Aj,0\rangle\rangle_{u}^{PF}\otimes
(|B0\rangle\rangle_{u}^{U(1)}+\eta|Bk\rangle\rangle_{u}^{U(1)}) \, .
\end{equation}
where $\eta=(-1)^{2a}$.
In deriving this expression one uses the field identification rule
$(j,n)\sim (k/2-j, k+n)$ and the following property of the matrix of
modular transformation~(\ref{smatr})
\begin{equation}
\label{prsymod}
S_{a,k/2-j}=(-1)^{2a}S_{aj} \, .
\end{equation}
To derive the symmetries preserved by the B-brane, one observes
from~(\ref{symmetr}) that a $Z_k^R$ invariant superposition of the
A-branes preserves only the current $J^3 + \bar{J}^3$ and breaks all
other currents; namely, any two $Z_k^R$ images only have this preserved
current in common. Performing further T-duality in the $\bar{J}^3$
direction flips the relative sign between the two terms in this
current and hence implies that the only current preserved by the
B-brane is
\begin{equation}
\label{nsymm}
J^3 - \bar{J}^3 = 0 \,  .
\end{equation}
We are now ready to apply this procedure to the brane (\ref{carsttw})
in order to generate symmetry breaking permutation branes.  Applying
the MMS procedure to one of the two $SU(2)$ groups will lead to brane~I,
while applying it to both groups will lead to the brane of type~II.
In this derivation we will need the following permuted Ishibashi
states in the~$U(1)_k\times U(1)_k$ and $PF_k\times PF_k$ theories.

The \emph{permuted} Ishibashi states satisfy the conditions 
\begin{equation}
\label{u(1)}
J_0^3\pm\bar{J}_1^3=0 \, , \quad  J_1^3\pm\bar{J}_0^3=0 \, .
\end{equation}
and  can be written as
 \begin{eqnarray}
\label{ishstg}
|r\rangle\rangle_{\tau\pm}^{U(1)_0\times U(1)_{\bar{1}}} &=& \exp\left[
\pm\sum_{n=1}^{\infty}{\alpha_{-n}^0\tilde{\alpha}^1_{-n}\over n}\right]
\sum_{l\in Z}|{r+2kl\over \sqrt{2k}}\rangle_0\otimes\overline{|\pm{r+2kl\over \sqrt{2k}}
\rangle}_1 \, \\
\label{ishstgg}
|r'\rangle\rangle_{\tau\pm}^{U(1)_1\times U(1)_{\bar{0}}} &=& \exp\left[
\pm\sum_{n=1}^{\infty}{\alpha_{-n}^1\tilde{\alpha}^0_{-n}\over n}\right]
\sum_{l'\in Z}
| \pm {r'+2kl'\over \sqrt{2k}}\rangle_1\otimes\overline{|{r'+2kl' \over \sqrt{2k}}
\rangle}_0 \, .
\end{eqnarray}
Here the signs in (\ref{u(1)}) are both taken to be plus or both minus.
Note that in  all these expressions there is a sum over the $l_0,l,l'$ labels, in
order to preserve the full extended~$U(1)_k$ symmetry algebra. For example,
all states formed from~(\ref{ishstg}) will preserve not only the first
symmetry in~(\ref{u(1)}) but also the conditions~\mbox{$\Gamma^{+}_{(0)} \pm
\bar{\Gamma}^{+}_{\bar{(1)}}= 0$} and \mbox{$\Gamma^-_{(0)} \pm
\bar{\Gamma}^-_{\bar{(1)}} = 0$}.

As for the abelian case, for the product of two parafermion theories
(${\cal A}_0^{PF(k)} \times {\cal A}_1^{PF(k)}$) we can define two
kinds of permuted states
\begin{eqnarray}
\label{ishfrtw}
|j_1,n_1\rangle\rangle_{\tau}^{PF_0\times PF_{\bar{1}}} &=&
\sum_{N}|j_1,n_1,N\rangle_0\otimes\overline{|j_1,n_1,N\rangle}_1 \, , \\
\label{ishsetw}
|j_2,n_2\rangle\rangle_{\tau}^{PF_1\times PF_{\bar{0}}} &=& 
\sum_{M}|j_2,n_2,M\rangle_1\otimes\overline{|j_2,n_2,M\rangle}_0 \, .
\end{eqnarray}
Here $j_i \in Z/2,\, n_i \in Z$ satisfy the constraint $2j + n = 0 \mod 2$ and 
an equivalence relation $(j,n) \sim (k/2-j, k+n)$.

\subsection{Boundary states for symmetry breaking type I branes}
\label{smbs}
We now want to construct the boundary state for the brane of type
I given in~(\ref{b5a}),
\begin{equation} 
\label{bb1}
(g_0,g_1)\Big|_{{\rm brane}} = \big(h_0f_0h_1^{-1}, h_1 f_1 h_0^{-1} e^{i \alpha {\sigma_3 \over 2}}\big) \,  .  
\end{equation}
Recall that, as we have derived before using the Langrangian approach,
this brane preserves the currents
\begin{eqnarray}
\label{bbc1}
J_0^3  &-&  \bar{J}_1^3=0 \, , \\
\label{bbc2}
\bar{J}_0^a &+&  J_1^a=0 \, , \quad (a=1,2,3) \, . 
\end{eqnarray}
To construct the boundary state, our starting point is the maximally
symmetric permutation brane~(\ref{carsttw}) which preserves the
symmetries~(\ref{preserv}). In order to reduce these symmetries down
to~(\ref{bbc1}), we will now show that one should apply the procedure
described in the previous section to the second~$SU(2)$ group in which
the permutation brane lives.  Namely, let us shift the
brane~(\ref{carsttw}) by multiplying it from the right with an element
$\omega_{(2)}^{l} = e^{{2\pi l i\over k}\sigma_3 }$ of the $Z_k^R$
subgroup of the second $SU(2)$ group. The shifted brane preserves the
symmetries
\begin{eqnarray}
\label{bbc11}
J_0^a  &+& \omega_{(2)}^l  \bar{J}_1^a \omega_{(2)}^{-l} =0 \, , \\
\label{bbc22}
\bar{J}_0^a &+&  J_1^a=0 \, , \hspace{2.5cm}  (a=1,2,3) \, ,
\end{eqnarray}
and is given by the Cardy state
\begin{equation}
\label{seccarom}
|a\rangle_{C}^{\omega^l_{(2)}}=\sum_{j}{S_{aj}\over S_{0j}}
|j\rangle\rangle_{\tau}^{SU(2)_1\times SU(2)_{\bar{0}}}\otimes |j\rangle\rangle_{\tau}^{\omega_{(2)}^l}\, , 
\end{equation}
where
\begin{equation} 
|j\rangle\rangle_{\tau}^{\omega_{(2)}^l}=\sum_N |j,N\rangle_0 \otimes (\omega_{(2)}^{l}\overline{|j,N\rangle}_{\bar{1}}) \, .
\end{equation}
As in the previous section, summing over the images and performing the
T-duality in the right sector,  will reduce the first
set of currents~(\ref{bbc11}) down to~(\ref{bbc1}), as desired.  As
far as the boundary state is concerned, summing over images will not
touch the first Ishibashi state in~(\ref{seccarom}) but will project the
second Ishibashi state  down to the~$Z_k^R$ invariant
components.  Using the permuted version of the
decomposition~(\ref{amax})
\begin{equation}
\label{perdec}
|j\rangle\rangle^{SU(2)_0\times SU(2)_{\bar{1}}}=\sum_{n=1}^{2k}{1+(-1)^{2j+n}\over 2}|j,n\rangle\rangle^{PF_0\times PF_{\bar{1}}}_{\tau}
\otimes |n\rangle\rangle^{U(1)_0\times U(1)_{\bar{1}}}_{\tau+}
\end{equation}
and applying T-duality to it, one obtains the
permuted B-type Ishibashi state of the initial $SU(2)$ theory,
\begin{multline}
\label{bpermis}
|B,j\rangle\rangle_{\tau}^{0\bar{1}} = {1+(-1)^{2j}\over 2}
|j,0\rangle\rangle_{\tau}^{PF_0\times PF_{\bar{1}}}\otimes|0\rangle\rangle_{\tau-}^{U(1)_0\times U(1)_{\bar{1}}}\\[1ex]
+{1+(-1)^{2j+k}\over 2}|j,k\rangle\rangle_{\tau}^{PF_0\times
 PF_{\bar{1}}}|k\rangle\rangle_{\tau-}^{U(1)_0\times U(1)_{\bar{1}}} \, .
\end{multline}
Here the permutation $U(1)$ and the permutation parafermion
Ishibashi states are given in formulas~(\ref{ishstg})
and~(\ref{ishfrtw}).  Using this expression the Cardy state for a
new brane can be written as:
\begin{equation}
\label{seccar}
|a\rangle_{C}^{(1)}=\sqrt{k}\sum_{j}{S_{aj}\over S_{0j}}
|j\rangle\rangle_{\tau}^{SU(2)_1\times SU(2)_{\bar{0}}}\otimes |B,j\rangle\rangle_{\tau}^{0\bar{1}} \, .
\end{equation}
Note also that since the boundary state~(\ref{seccar}) is ``derived''
from the maximally symmetric boundary state~(\ref{carsttw}), it is
characterised with a \emph{single} primary~$j$ as was the case for
the brane~(\ref{carsttw}). This is again related to the fact that in
the effective description~(\ref{bb1}), there is only one independent
parameter~($f\equiv f_0 f_1$).\footnote{This can be easily be seen by
changing coordinates as $h_0 \rightarrow h_0 f_1^{-1}$.}

To check the consistency of the proposed boundary state, one should
check, as usual, the Cardy condition. Since we are in a theory which
admits several different types of branes, one should in principle
check these conditions for the type~I brane with any of the other
branes in the spectrum. We have done the calculation involving two
branes of type~I, with one of type~I and a permutational brane, and
with a brane which is direct product of two $SU(2)$ A-branes.  The tree-level amplitude
between two Cardy states for two type~I branes reduces, after the S-modular
transformation reduces, to
\begin{equation}
\label{pf1}
Z_{a_1a_2}=\sum_{r,j',j''}\sum_{n_1,n_2} N^r_{a_1a_2}N^{j'}_{rj''}\chi_{j'}(q)\chi_{j'',n_1}(q)\psi_{n_2}(q){1+ (-1)^{n_1+n_2}\over 2} \, , 
\end{equation}
hence satisfying the Cardy
requirement.  The annulus amplitude between the type~I and the maximally symmetric
permutation brane~(\ref{carsttw}) reduces, after the
S-modular transformation, to
\begin{equation}
\label{pf2}
Z_{a_1a_2}=\sum_{r,j',j''}\sum_{n_1} N^r_{a_1a_2}N^{j'}_{rj''}\chi_{j'}(q)\chi_{j'',n_1}(q){q^{1/48}\over \prod_m(1-q^{m-1/2})} \, , 
\end{equation} 
while the one between states of type~I and the brane~(\ref{carst}) reduces to
\begin{equation}
\label{pf3}
Z_{a,(a_0a_1)}=
\sum_{r,j'}\sum_n N^r_{a_0a_1}N^{j'}_{ra}\chi_{j',n}(q^{1/2}){(q^{1/2})^{1/48}\over \prod_m(1-(q^{1/2})^{m-1/2})}\, .
\end{equation}
Here the factor ${q^{1/48}\over \prod_m(1-q^{m-1/2})}$ is the
partition function of a scalar with mixed Neumann-Dirichlet type
boundary conditions.
The details of calculations of (\ref{pf1}), (\ref{pf2}) and (\ref{pf3})
can be found in the appendix B.

We will now show that the boundary state (\ref{seccar}) reproduces the
effective brane geometry~(\ref{thetprun}).  In the large~$k$ limit the
second term in~(\ref{bpermis}) can be ignored.  As in section
\ref{sympres} one should compute the overlap
$\langle\vec{\theta}_0,\vec{\theta}_1|a\rangle_{C}^{(1)}$.  We will
again use the formula~(\ref{ishov}), but taking into account that the
matrix ${\cal D}^{(1)}$ derived for the first group has left index $0$
and the right index $m$, whereas the ${\cal D}^{(2)}$ matrix derived
for the second group has left index $m$ and the right index
$0$. Therefore, the overlap is again given by formula~(\ref{twov}),
but with~$n$ set to zero.  Using furthermore~(\ref{matprod}) we arrive
at the equation
\begin{equation}
\label{seccarov}
\langle\vec{\theta}_0,\vec{\theta}_1|a\rangle_{C}^{(1)}
\sim
\sum_{j}{k^{3/2}\over \pi}\sin[(2j+1)\hat{\psi}]\, 
{\cal D}^j_{00}(g_0(\vec{\theta}_0)g_1(\vec{\theta}_1))
\end{equation}
Next we will need the relation between the Wigner D-functions and the
Legendre polynomials $ P_j(\cos\tilde{\theta})$ given by ${\cal
D}^j_{00}=P_j(\cos\tilde{\theta})$, as well as the formula for the
generating function for Legendre polynomials
\begin{equation}
\label{genleg}
\sum_n t^n P_n(x)={1\over \sqrt{1-2tx+t^2}}\, .
\end{equation}
Using these expressions equation (\ref{seccarov}) can be simplified to
\begin{equation}
\langle\vec{\theta}_0,\vec{\theta}_1|a\rangle_{C}^{(1)}
\sim
{\Theta(\cos\hat{\tilde{\theta}}-\cos2\hat{\psi})\over
\sqrt{\cos\hat{\tilde{\theta}}-\cos2\hat{\psi}}} \, , 
\end{equation}
where $\Theta$ is the step function.  This indeed coincides with the
expression for the effective geometry~(\ref{thetprun}).

\subsection{Boundary states for symmetry breaking type II branes}
\label{brtwo}
Let us now turn to the type II brane~(\ref{b5b})
\begin{equation} 
\label{br5}
(g_0,g_1)\Big|_{{\rm brane}} = \big(h_0f_0h_1^{-1} e^{i \beta {\sigma_3 \over 2}}, h_1 f_1 h_0^{-1} e^{i \alpha {\sigma_3 \over 2}}\big) \,  ,  
\end{equation}
which  preserves the currents 
\begin{equation}
\label{IIcr}
J_0^3  - \bar{J}_1^3=0 \, , \quad \bar{J}_0^3  -   J_1^3=0 \, . 
\end{equation}
This brane has a structure which is very similar to the type~I
brane. It can be derived from this brane by applying the described
procedure (with right cosetting) to the \emph{first} $SU(2)$ group in
which brane~I lives.  As for brane~I, this procedure will reduce the
currents~(\ref{bbc1}) and~(\ref{bbc2}) down to~(\ref{IIcr}). At the
level of the boundary state, the Ishibashi state in the $0\bar{1}$
sector will remain unchanged, while the $\bar{0}1$ Ishibashi
state~(\ref{permish1}) will be projected down to a~$Z_{k,1}^R$ invariant
state (where subscript 1, indicates that this action is taken in the first $SU(2)$ group). Finally, applying the T-duality 
we obtain the boundary state
\begin{eqnarray}
\label{thircar}
|a\rangle_{C}^{(2)}&=& k\sum_{j}{S_{aj}\over S_{0j}}
|B,j\rangle\rangle_{\tau}^{0\bar{1}}\otimes|B,j\rangle\rangle_{\tau}^{1\bar{0}}
\end{eqnarray}
where $|B,j\rangle\rangle_{\tau}^{1\bar{0}}$ is defined as
in~(\ref{bpermis}) with~$0$ and~$1$ exchanged,
and the coefficients in the linear combination are fixed by the Cardy
condition.  For even~$k$ the tree-level amplitude between the
states~(\ref{thircar}) reduces to
\begin{multline}
\label{pff1}
Z_{a_1a_2}=\sum_{r,j',j''}\sum_{n_1,n_2}\sum_{n_3,n_4} N^r_{a_1a_2}N^{j'}_{rj''}
\chi_{j',n_1}(q)\chi_{j'',n_3}(q)\psi_{n_2}(q)\psi_{n_4}(q)\\
\times {(1+(-1)^{n_1+n_2})(1+(-1)^{n_3+n_4})\over 4}\,.
\end{multline}
For an odd $k$ (\ref{thircar}) can be simplified and written as
\begin{multline}
\label{thircarodd}
|a\rangle_{C}^{(2)}=\\
\begin{aligned}
k \sum_{j}{S_{aj}\over S_{0j}}\bigg[ 
{1+(-1)^{2j}\over 2}
|j,0\rangle\rangle_{\tau}^{PF_0\times PF_{\bar{1}}}&\otimes|j,0\rangle\rangle_{\tau}^{PF_1\times PF_{\bar{0}}}
\otimes|0\rangle\rangle_{\tau-}^{U(1)_0\times U(1)_{\bar{1}}}\otimes|0\rangle\rangle_{\tau-}^{U(1)_1\times U(1)_{\bar{0}}}\\[1ex]
+{1+(-1)^{2j+k}\over 2}
|j,k\rangle\rangle_{\tau}^{PF_0\times PF_{\bar{1}}}&\otimes|j,k\rangle\rangle_{\tau}^{PF_1\times PF_{\bar{0}}}
\otimes|k\rangle\rangle_{\tau-}^{U(1)_0\times
  U(1)_{\bar{1}}}\otimes|k\rangle\rangle_{\tau-}^{U(1)_1\times
  U(1)_{\bar{0}}}\bigg]\,.
\end{aligned}
\end{multline}
The tree level amplitude between the states~(\ref{thircarodd}) is
\begin{equation}
\label{pff2}
Z_{a_1a_2}=\sum_{r,j',j''}\sum_{n_1,n_2}\sum_{n_3,n_4} N^r_{a_1a_2}N^{j'}_{rj''}
\chi_{j',n_1}(q)\chi_{j'',n_3}(q)\psi_{n_2}(q)\psi_{n_4}(q){1+\eta(-1)^{n_2+n_4}\over 4}
\end{equation}
where $\eta=(-1)^{2a_1+2a_2}$. The  Cardy condition is satisfied in this case  
because, after taking into account field identification in the parafermionic sector, one can show that  each state in the sum appears twice, and therefore all states appear with integer coefficient.
The calculations leading to (\ref{pff1}) and (\ref{pff2}) follow closely to that of for
(\ref{pf1}) outlined in the appendix B.

To derive the effective geometry of this brane, one follows the same
arguments we as presented for the type~I brane. The overlap with the
localised bulk probe is given by
\begin{equation}
\label{thircarov}
\langle\vec{\theta}_0,\vec{\theta}_1|a\rangle_{C}^{(2)}
\sim
\sum_{j}{k^2\over \pi}\sin[(2j+1)\hat{\psi}]
P_j(\cos\tilde{\theta}_0)P_j(\cos\tilde{\theta}_1) \, .
\end{equation}
Using now the formula~\cite{Vilenkin:1968nk}
\begin{equation}
\label{prodlen}
P_j(\cos\tilde{\theta}_0)P_j(\cos\tilde{\theta}_1)=
{1\over \pi}\int_{|\tilde{\theta}_0-\tilde{\theta}_1|}^{\tilde{\theta}_0+\tilde{\theta}_1}
P_j(\cos\theta){\sin\theta d\theta\over
\sqrt{[\cos\theta-\cos(\tilde{\theta}_0+\tilde{\theta}_1)][\cos(\tilde{\theta}_0-\tilde{\theta}_1)-\cos\theta]}}
\end{equation}
and equation~(\ref{genleg}) we obtain
\begin{equation}
\label{thirfin}
\langle\vec{\theta}_0,\vec{\theta}_1|a\rangle_{C}^{(2)}
\sim
{1\over \pi}\int_{|\tilde{\theta}_0-\tilde{\theta}_1|}^{\tilde{\theta}_0+\tilde{\theta}_1}
{\Theta(\cos\theta-\cos2\hat{\psi})\over
\sqrt{\cos\theta-\cos2\hat{\psi}}}  {\sin\theta d\theta\over
\sqrt{[\cos\theta-\cos(\tilde{\theta}_0+\tilde{\theta}_1)][\cos(\tilde{\theta}_0-\tilde{\theta}_1)-\cos\theta]}} \, .
\end{equation}
The integral~(\ref{thirfin}) is different from zero if
$\cos(\tilde{\theta}_0-\tilde{\theta}_1)\geq
\cos2(\hat{\psi}_0+\hat{\psi}_1)$ which is precisely the
condition~(\ref{thirbr}).

\subsection{Boundary states for symmetry breaking type III branes}
\label{bsiii}

Finally, let us consider the type III brane given in~(\ref{nnbrd}),
\begin{equation} 
\label{scunp}
(g_0,g_1)\Big|_{{\rm brane}} = \big(h_0f_0h_1^{-1} e^{i \alpha {\sigma_3 \over 2}}, h_1 f_1 h_0^{-1} e^{-i \alpha {\sigma_3 \over 2}}\big) \,  ,  
\end{equation}
with preserved currents
\begin{equation}
\label{IIIcr}
J_0^3  +\bar{J}_0^3=0 \, , \quad J_1^3 +  \bar{J}_1^3=0 \, . 
\end{equation}
This brane has a structure which is different from the two previous
branes. To derive the boundary state for this brane we start from the
maximally symmetric permutation brane~(\ref{carsttw}) and decompose
the factors~(\ref{permish1}) and~(\ref{permish2}) according to 
the equation (\ref{perdec}) and its $0$ and $1$ exchanged version.
We then apply the permutational symmetry between the two $U(1)$
factors in the~$0\bar{1}$ and~$1 \bar{0}$ subsectors to unpermute
them. 
Since the final state has to be~$Z_k$ invariant (in accordance with the
decomposition~(\ref{decom})),\footnote{Note that the action of the
$Z_k$ group in~(\ref{decom}) is different from the one used in the
construction of the Lens space $Z_k^R$. The action of $\omega \in Z_k$
on an arbitrary state $|j,n\rangle^{PF} |m \rangle^{U(1)}$ is given by
\begin{equation}
\omega |j,n\rangle^{PF} |m \rangle^{U(1)} = \omega^{n-m}
|j,n\rangle^{PF} |m \rangle^{U(1)} \, .
\end{equation}
} one is forced to restrict
to the subsector with~$n=m$.
The boundary state for the brane~(\ref{scunp}) can hence be written as
\begin{multline}
\label{thircar2}
|a\rangle_{C}^{(3)}=\sqrt{2k}\sum_{j}\sum_n{S_{aj}\over S_{0j}}{1+(-1)^{2j+n}\over 2}\\[1ex]
|j,n\rangle\rangle_{\tau}^{PF_0\times PF_{\bar{1}}}\otimes|j,n\rangle\rangle_{\tau}^{PF_1\times PF_{\bar{0}}}\otimes
|A,n\rangle\rangle_{u}^{U(1)_0}\otimes|A,n\rangle\rangle_{u}^{U(1)_1} \, .
\end{multline}
As before, to check the consistency of the boundary state, one needs
to ensure that the Cardy condition holds.  The tree-level amplitude
between two type III boundary states reduces, after the S-modular
transformation, to
\begin{equation}
\label{pfff1}
Z_{a_1a_2}=\sum_{r,j',j''}\sum_{n_1}\sum_{n_2,n_3} N^r_{a_1a_2}N^{j'}_{rj''}
\chi_{j',n_1+n_2+n_3}(q)\chi_{j'',n_1}(q)\psi_{n_2}(q)\psi_{n_3}(q) \, ,
\end{equation}
and hence is consistent with the Cardy condition.  Similarly, the
annulus amplitude between brane~III~(\ref{thircar2}) and the
maximally symmetric permutational brane~(\ref{carsttw}) is
\begin{equation}
\label{pfff2}
Z_{a_1a_2}=\sum_{r,j',j''}\sum_{n_1,n_2} N^r_{a_1a_2}N^{j'}_{rj''}
\chi_{j',n_1+n_2}(q)\chi_{j'',n_1}(q)\psi_{n_2}(q^{1/2}) \, .
\end{equation}
and between type III and (\ref{carst})
\begin{equation}
\label{pfff3}
Z_{a,(a_0a_1)}=
\sum_{j',r}\sum_{n_1,n_2} N^r_{a_0a_1}N^{j'}_{ra}\chi_{j',n_1+n_2}(q^{1/2})\psi_{n_1}(q)\psi_{n_2}(q)\, .
\end{equation} 
and type III and type I
\begin{equation}
\label{pfff4}
Z_{a_1a_2}=
\sum_{r,j',j''}\sum_{n,m}N^r_{a_1a_2}N^{j'}_{rj''}\chi_{j',n}(q)\chi_{j'',m}(q){(q^{1/2})^{1/48}\over \prod_m(1-(q^{1/2})^{m-1/2})}\, .
\end{equation}
The details of calculations (\ref{pfff1})-(\ref{pfff4}) are provided in the appendix B.
 
Now we calculate the effective geometry corresponding to this
boundary state. As before, to obtain the effective geometry, one
should compute the overlap
$\langle\vec{\theta}_0,\vec{\theta}_1|a\rangle_{C}^{(3)}$.  In the
large-$k$ limit the overlap reduces to
\begin{equation}
\langle\vec{\theta}_0,\vec{\theta}_1|a\rangle_{C}^{(3)}\sim
\sum_j\sum_n\sin(2j+1){\cal D}^j_{nn}(g_0(\vec{\theta}_0)){\cal D}^j_{nn}(g_1(\vec{\theta}_1)) \, .
\end{equation}
One needs to use the fact that Wigner D-functions may be represented as
a product of three functions, each of which depends only on one Euler
coordinate,
\begin{equation}
{\cal D}^j_{nm}(g(\vec{\theta}))=e^{-i(n\chi+m\varphi)}d^j_{nm}(\cos\tilde{\theta}) \, , 
\end{equation}
where $d_{nm}^j$ are real functions satisfying the relation (note that
there is no summation assumed for the repeated indices)
\begin{equation}
d^j_{nn}(\cos\tilde{\theta}_0)d^j_{nn}(\cos\tilde{\theta}_1)={1\over 2\pi}\int_{-\pi}^{\pi}
e^{in(\gamma-\xi)}d^j_{nn}(\cos\hat{\Theta})d\gamma \, , 
\end{equation}
The functions $\hat{\Theta}$ and $\xi$ are functions of
$\tilde{\theta}_0,\tilde{\theta}_1$ and $\gamma$ defined in
equations~(\ref{thhh}) and (\ref{phiprrr}). The overlap of the
boundary state with the bulk probe can be written as
\begin{eqnarray}
\langle\vec{\theta}_0,\vec{\theta}_1|a\rangle_{C}^{(3)} &\sim&
\sum_j\sum_n\int_{-\pi}^{\pi}\sin(2j+1)e^{in(\gamma-\xi-2\tilde{\phi}_0-2\tilde{\phi}_1)}d^j_{nn}(\cos\hat{\Theta})d\gamma \, \nonumber \\
&\sim&
\sum_j\sum_n\int_{-\pi}^{\pi}\sin(2j+1)D^j_{nn}(\hat{\Theta},\gamma/2-\xi/2-\tilde{\phi}_0-\tilde{\phi}_1,\gamma/2-\xi/2-\tilde{\phi}_0-\tilde{\phi}_1 )d\gamma \, \nonumber \\
\end{eqnarray}
Now repeating the same steps as in (\ref{twovtr}) and (\ref{fresu}) we get
\begin{equation}
\langle\vec{\theta}_0,\vec{\theta}_1|a\rangle_{C}^{(3)}\sim
\int_{-\pi}^{\pi}{\delta(\psi-\hat{\psi})\over \sin\hat{\psi}}d\gamma \, , 
\end{equation}
where
\begin{equation}
\cos\psi=\cos{\hat{\Theta}\over 2}
\cos(\gamma/2-\xi/2-\tilde{\phi}_0-\tilde{\phi}_1)
\end{equation}
From this equation it follows that the brane consist of all those
points for which the expression in the argument of the~$\delta$
function has a root for~$\gamma$. This is the same condition as the
one coming from equation~(\ref{baseq1}), obtained in the Langrangian
approach.

\section{Discussion}

In this paper we have presented several new types of symmetry breaking
branes.  While most of our analysis was focused on exploring different
ways in which one can break part of the diagonal affine algebra by
branes, the understanding of the properties of these branes was only
briefly touched upon. In particular, issues concerning the embedding
of these branes in supersymmetric models, their stability, a
comparison of the spectra between the CFT and the effective approaches
and so forth should definitely still be explored further.  A
preliminary investigation of the supersymmetry of these branes using
the probe kappa-symmetry approach indicates that most of the branes
are non-supersymmetric. Only a special class of maximally symmetric,
permutation branes seems to preserve a fraction of
supersymmetry. While most of the required effective analysis as found
in~\cite{Bain:2002nq} or~\cite{Bain:2002tq} is complicated when
applied directly to a target that is a product of groups,
simplifications might occur in the Penrose limit applied to branes
along the diagonal geodesic which winds between the groups, as
in~\cite{Sarkissian:2003jn}.  We hope to address some of these issues
in the future.

\section*{Acknowledgements}
We would like to thank Pedro Bordalo, Annamaria Font, Kasper Peeters, Sakura
Sch\"afer-Nameki and Volker Schomerus for useful discussions.

\vfill
\eject
\appendix

\section{Some details of the calculations}
\subsection{Symmetries of the brane I}
\label{invariance}
In this appendix we present a detailed calculation of the invariance of
the string action~(\ref{actwzw}) for the brane~(\ref{genbr}) under the
symmetry transformations~1 and~2 of section~\ref{gr1}.  Under
the first transformation the change of the bulk action is~(\ref{delreac})
with $g_0=K_0$ and $g_1=K_1L$, while the change of the first term in~(\ref{ngom}) is
\begin{equation}
\label{chwz}
\omega^{(2)}(h_0,kh_1)-\omega^{(2)}(h_0,h_1)=-\Tr\Big(k^{-1}{\rm
  d}k(K_0^{-1}{\rm d}K_0+{\rm d}K_1K_1^{-1})\Big) \, .
\end{equation}
The  change in the second term in~(\ref{ngom}) is given by
\begin{equation}
\label{chwzp}
\Delta \Big(- \Tr\big(K_1^{-1}{\rm d}K_1{\rm d}LL^{-1}\big)\Big)=
-\Tr\big(k^{-1}{\rm d}k(K_1{\rm d}LL^{-1}K_1^{-1})\big) \, .
\end{equation}
Hence putting all terms together, one gets that the change of
$\omega^{(2)}(h_0,h_1,L)$ exactly cancels~(\ref{delreac}).

Under the second transformation in~(\ref{gr1}) the change in the bulk
term in the action is
\begin{equation}
\label{chac1}
\Delta S = -{k\over 4\pi}\int_D\Tr\Big(k^{-1}{\rm d}k\big({\rm
  d}K_0K_0^{-1}+K_1^{-1}{\rm d}K_1+L^{-1}dL\big)\Big)
\end{equation}
where we used that $[k,L]=0$.  The  change in the first term in (\ref{ngom}) is given by
\begin{equation}
\label{chwz1}
\omega^{(2)}(kh_0,h_1)-\omega^{(2)}(h_0,h_1)=-\Tr\Big(k^{-1}{\rm
  d}k\big({\rm d}K_0K_0^{-1}+K_1^{-1}{\rm d}K_1\big)\Big)
\end{equation}
and in the change in the second term by
\begin{equation}
\label{chwzp1}
\Delta \Big(- \Tr\big(K_1^{-1}{\rm d}K_1{\rm
  d}LL^{-1}\big)\Big)=\Tr\big(k^{-1}{\rm d}k{\rm d}LL^{-1}\big)\,,
\end{equation}
again due to $[k,L]=0$. Hence the total change of the action is,
\begin{equation}
\label{chf1}
\Delta \left(S-{k\over 4\pi}\int_D\omega^{(2)}(h_0,h_1,L)\right)=
{k\over 2\pi}\int_D\Tr\big(k^{-1}{\rm d}k{\rm d}LL^{-1}\big) \, ,
\end{equation}
which vanishes for all $[k,L]=0$.

The change in the bulk part of the action under the transformations 3' is 
\begin{equation}
\label{chac2}
\Delta S = -{k\over 4\pi}\int_D \Tr\Big(k^{-1}{\rm d}k\big({\rm
  d}K_0K_0^{-1}-K_1^{-1}{\rm d}K_1-L^{-1}dL\big)\Big) \, , 
\end{equation}
while the  change in the second term in~(\ref{ngom}) is
\begin{equation}
\label{chwzp2}
\Delta \Big(- \Tr\big(K_1^{-1}{\rm d}K_1{\rm d}LL^{-1}\big)\Big)
=\Tr\big(k^{-1}{\rm d}k(2K_1^{-1}{\rm d}K_1+{\rm d}LL^{-1}\big)
\end{equation}
which altogether with~(\ref{chwz1}) leads to a vanishing total
variation of the action.

\subsection{Symmetries of the brane III}
\label{symm}
In this section we prove invariance of the brane~(\ref{genbrrd})
under the transformations 3'' and 4'' in section~\ref{gen}. The
gauge-invariant two-form for this brane is given by
\begin{equation}
\label{omtl}
\omega^{\rm WZ}(g)\Big|_{\rm brane}={\rm d}\Big{(}\omega^{(2)}(h_0,h_1)-{\rm
Tr}\big(K_0^{-1}{\rm d}K_0{\rm d}LL^{-1}-K_1^{-1}{\rm d}K_1{\rm d}LL^{-1}\big) \Big{)} \, .
\end{equation}
Under the  transformation 3'' the bulk action changes by the amount
\begin{equation}
\label{bacch}
\begin{aligned}
\Delta S&={k\over 4\pi}\int_D{\rm Tr}\Big(k^{-1}{\rm d}k\big(g_0k^{-1}{\rm d}kg_0^{-1}-g_0^{-1}{\rm d}g_0-{\rm d}g_0g_0^{-1}\big)\Big)\\
&={k\over 4\pi}\int_D{\rm Tr}\Big(k^{-1}{\rm d}k\big(K_0k^{-1}{\rm d}kK_0^{-1}-K_0^{-1}{\rm d}K_0-{\rm d}K_0K_0^{-1}-L^{-1}{\rm d}L\\
&\hskip.6\textwidth -K_0{\rm d}LL^{-1}K_0^{-1}\big)\Big)\,.
\end{aligned}
\end{equation}
The change of the first term in the two-form~(\ref{omtl}) is given
by~(\ref{chwz1}) while the change of the second term is
\begin{multline}
\label{chpard}
\Delta\Big(-{\rm Tr}\big(K_0^{-1}{\rm d}K_0{\rm d}LL^{-1}-K_1^{-1}{\rm d}K_1{\rm d}LL^{-1}\big)\Big)=\\
{\rm Tr}\Big(k^{-1}{\rm d}k\big(K_1^{-1}{\rm d}K_1-K_0^{-1}{\rm d}K_0-{\rm d}LL^{-1}
+K_0{\rm d}kk^{-1}K_0^{-1}-K_0{\rm d}LL^{-1}K_0^{-1}\big)\Big) \, .
\end{multline}
Collecting all terms together, we see that the action is invariant
under the transformation~3''. Invariance of the action under
the transformations~4'' is proved in the same way.


\section{Proof of the Cardy conditions}
\label{cardyproof}
One of the necessary consistency conditions which any boundary state
has to satisfy is the Cardy condition: for any pair of boundary states
$|\alpha \rangle $, $|\beta \rangle$ in the theory, the closed string
tree-level amplitude between them should, after a modular
transformation, be expressible as the partition function of a CFT on a
strip with boundary conditions~$\alpha$ and~$\beta$ at each end of the
strip.  The requirement that the boundary state satisfies the Cardy
condition restricts the coefficients in the Cardy state as follows.
We present here the details of the calculation for~(\ref{carsttw})
following~\cite{Recknagel:2002qq}. The computation for other branes in
section~\ref{smbs} is similar, so only the results are presented in
the main text.

Let us consider the following tree amplitudes between the two
permutational branes given in~(\ref{carsttw}) and between the
permutational and factorisable branes of~(\ref{sucar}),
\begin{eqnarray}
\label{TC1}
Z_{a_1,a_2}&=&{}_{{\cal P}}\langle a_1|\exp(-\pi iH_{(cl)}/T)|a_2\rangle_{{\cal P}} =
{}_{{\cal P}}\langle
a_1|(\tilde{q}^{1/2})^{L_{0}+\overline{L_{0}}-c/12}|a_2\rangle_{{\cal
P}} \, , \\
\label{TC2}
Z_{a,(a_0a_1)}&=&{}_{{\cal P}}\langle a|\exp(-\pi iH_{(cl)}/T)|a_0,a_1\rangle =
{}_{{\cal P}}\langle a|(\tilde{q}^{1/2})^{L_{0}+\overline{L_{0}}-c/12}|a_0,a_1\rangle \, .
\end{eqnarray}
Here $\tilde{q}\equiv e^{-2\pi i/T}$ and $L_0=L_0^{(0)}+L_0^{(1)}$,
$\bar{L}_0=\bar{L}_0^{(0)}+\bar{L}_0^{(1)}$. We want to show that
after the S-modular transformation, these expressions can be written
as
\begin{equation}
\label{LC}
Z_{\alpha\beta}=\sum_{i}n^{i}_{\alpha\beta}\chi_{i}(q) \, .
\end{equation}
Here $\alpha$ and $\beta$ are conditions defined by the states
(\ref{carst}) and (\ref{carsttw}), $q\equiv \exp(-2\pi iT)$ is the
open string modular parameter, $n^{i}_{\alpha\beta}$ are some positive
\emph{integer} numbers, and $\chi_{i}(q)$ are characters for
\emph{some} conformal symmetry algebra. Note that, depending on the
choice of the boundary conditions, the same initial bulk algebra leads
to different open string conformal algebras.

The first expression (\ref{TC1})  becomes 
\begin{equation}
\label{ishover}
\begin{aligned}
{}_{\cal P}\langle\langle j_1,j_1|&(\tilde{q}^{1/2})^{L_{0}+\overline{L}_{0}-c/12}|j_2,j_2 \rangle \rangle_{\cal P} =\\[1.5ex]
&=\sum_{M,N,M'N'}\tilde{q}^{h(j_1,M)+h(j_2,N)-c/12}
\begin{aligned}[t]
 &\big({}_0\langle j_1,M|j_2,M'\rangle_0\big)\big({}_1\overline{\langle j_1,M|j_2,M'\rangle_1}\big)\\[1ex]
 &\times\big({}_0\overline{\langle j_1,N|j_2,N'\rangle_0}\big) \big({}_1\langle j_1,N|j_2,N'\rangle_1\big)
\end{aligned}\\[1.5ex]
&=\sum_{M,N}\tilde{q}^{h(j_1,M)+h(j_2,N)-c/12}\delta_{j_1,j_2}=\chi_{j_1}
(\tilde{q})\chi_{j_2}(\tilde{q})\delta_{j_1,j_2} \, .
\end{aligned}
\end{equation}

Applying the modular transformations to this expression, and using the
symmetry properties of the S-matrix ($S_{ij} = S_{ji}$) as well as the
Verlinde formula, one obtains\footnote{Recall that the Verlinde
formula is given by
\begin{equation}
N_{ij}^k = \sum_{p} {S_{i p} S_{j p} \bar{S}_{p k} \over S_{0p}} \, ,
\end{equation}
where $\bar{S} = C S$ is a complex conjugated S-matrix, $N_{ij}^k$ are
fusion matrices and the 0~index in~S denotes the vacuum representation.
In the case of a real S-matrix, this
can be rewritten as 
\begin{equation}
\label{verfor}
{S_{aj}S_{lj}\over S_{0j}}=\sum_k N^k_{al}S_{kj} \, .
\end{equation}  
}
\begin{equation}
\label{part}
\begin{aligned}
Z_{a_1,a_2}&=\sum_{j,k,l}{S_{a_1j}\over S_{0j}}{S_{a_2j}\over S_{0j}}S_{jk}S_{jl}\chi_{k}(q)\chi_{l}(q) \\[1ex]
&= \sum_{r,k,l} N^r_{a_1a_2}N^k_{rl}\chi_{k}(q)\chi_{l}(q) \, .
\end{aligned}
\end{equation}
To put this expression in the form~(\ref{LC}) we need to realise that
the product of two characters in~(\ref{part}) corresponds to a
\emph{single} character of the total~$SU(2)\times SU(2)$ group, since
the primaries of the total group are labeled by two (rather then one
number). Hence we can write $\chi_k(q) \chi_l(q) \equiv
\chi_{(k,l)}(q)$. Also the sum of the two fusion matrices
in~(\ref{part}) is obviously defining a set of positive integer
numbers, which we can denote with~$N_{a_0 a_1}^{k,l}$. Using all these
ingredients we see that~(\ref{part}) is of the required form.

It is instructive to compare the partition function we have just
computed with the the partition function between the boundary states
for the direct product of two~$SU(2)$ branes
\begin{equation}
\label{twof}
Z_{(a_0,a_1),(a_0',a_1')}= \sum_{r,k,l} N^r_{a_0 a_0'}N^k_{a_1 a_1'}\chi_{r}(q)\chi_{l}(q) \, .
\end{equation}
We see that the difference with respect to the~(\ref{part}) is
contained in the form of the fusion coefficients; while in the
formula~(\ref{part}) the coefficients appear as product of fusion
matrices ($(N_i)_{jk} = N_{ij}^k $), they appear in the
expression~(\ref{twof}) in an ``uncoupled'' form.

Next we turn to the computation of (\ref{TC2}).
\begin{equation}
\label{secalo}
Z_{a,(a_0a_1)}=\sum_{j,j_1,j_2}{S_{aj}\over S_{0j}}{S_{a_0j_0}S_{a_1j_1}\over \sqrt{S_{0j_0}S_{0j_1}}}
{}_{\cal P}\langle j,j|(\tilde{q}^{1/2})^{L_{0}+\overline{L_{0}}-c/12}|j_0\rangle\rangle_u^{SU(2)_0}
|j_1\rangle\rangle_u^{SU(2)_1} \, .
\end{equation}
In this case the overlap between the Ishibashi states is
\begin{equation}
\label{ishot}
\begin{aligned}
{}_{\cal P}\langle j,j|&(\tilde{q}^{1/2})^{L_{0}+\overline{L_{0}}-c/12}|j_0\rangle\rangle_u^{SU(2)_0}
|j_1\rangle\rangle_u^{SU(2)_1}=\\[1.5ex]
&= \sum_{M,N,M'N'}\tilde{q}^{h(j_1,M)+h(j_2,N)-c/12}
 \begin{aligned}[t]
 &\big({}_0\langle j,M|j_0,M'\rangle_0\big)\big({}_1\overline{\langle j,M|j_1,N'\rangle_1}\big)\\[1ex]
 &\times\big({}_0 \overline{\langle j,N|j_0,M'\rangle_0}\big)\big({}_1\langle j,N|j_1,N'\rangle_1\big)
 \end{aligned}\\[1.5ex]
&=
\sum_{M}\tilde{q}^{2h(j,M)-c/12}\delta_{j,j_0}\delta_{j,j_1}=\chi_{j}(\tilde{q}^2)\delta_{j,j_0}\delta_{j,j_1}
\, .
\end{aligned}
\end{equation}
After the modular transformations and manipulations similar to those
we have already used, the partition function becomes
\begin{equation}
Z_{a,(a_0a_1)}=\sum_{j,k,r}N^r_{a_0a_1}S_{rj}{S_{aj}\over S_{0j}}S_{jk}\chi_{k}(q^{1/2})=
\sum_{k,r}N^r_{a_0a_1}N^k_{ra}\chi_{k}(q^{1/2})\, .
\end{equation}
We see that the partition function is expressible as a sum over
characters in a twisted sector of a theory orbifolded by the
permutation symmetry~\cite{Klemm:1990df,Borisov:1998nc}.

For boundary states (\ref{seccar}) we have
\begin{equation}
\begin{aligned}
{}_{C}^{(1)}\langle a_1|&(\tilde{q}^{1/2})^{L_{0}+\overline{L_{0}}-c/12}|a_2\rangle_{C}^{(1)}=\\[1.5ex]
&k\sum_{j}{S_{a_1j}S_{a_2j}\over S_{0j}S_{0j}}
\left({1+(-1)^{2j}\over 2}\chi_j(\tilde{q})\chi_{j0}(\tilde{q})\psi_0(\tilde{q})+
{1+(-1)^{2j+k}\over 2}\chi_j(\tilde{q})\chi_{jk}(\tilde{q})\psi_k(\tilde{q})\right)
\end{aligned}
\end{equation}
which using the Verlinde formula can be written as
\begin{equation}
\begin{aligned}
{}_{C}^{(1)}\langle a_1|&(\tilde{q}^{1/2})^{L_{0}+\overline{L_{0}}-c/12}|a_2\rangle_{C}^{(1)}= \\[1.5ex]
&k\sum_{j,r}N^r_{a_1,a_2}{S_{rj}\over S_{0j}}
\left({1+(-1)^{2j}\over 2}\chi_j(\tilde{q})\chi_{j0}(\tilde{q})\psi_0(\tilde{q})+
{1+(-1)^{2j+k}\over 2}\chi_j(\tilde{q})\chi_{jk}(\tilde{q})\psi_k(\tilde{q})\right)
\end{aligned}
\end{equation}
Now we will evaluate all four terms in parenthesis after modular transformation
\begin{equation}
\label{parsu}
\begin{aligned}
{k\over 2}&\sum_{j,r}N^r_{a_1,a_2}{S_{rj}\over S_{0j}}\chi_j(\tilde{q})\chi_{j0}(\tilde{q})\psi_0(\tilde{q})\\
&={k\over 2}\sum_{j,r}N^r_{a_1,a_2}\sum_{j',j'',n_1,n_2}{S_{rj}S_{jj'}S_{jj''}\over S_{0j}}
\chi_{j'}(q){\chi_{j''n_1}(q)\psi_{n_2}(q)\over \sqrt{2k}\sqrt{2k}}\\
&={1\over 4}\sum_{r}\sum_{j',j'',n_1,n_2}N^r_{a_1,a_2}N^{j'}_{r,j''}\chi_{j'}(q)\chi_{j''n_1}(q)\psi_{n_2}(q)
\end{aligned}
\end{equation}
Here we used matrix of modular transformation of $U(1)_k$ theory given by formula (\ref{modtr}).
\begin{equation}
\label{parsu2}
\begin{aligned}
{k\over 2}&\sum_{j,r}N^r_{a_1,a_2}{S_{rj}\over S_{0j}}(-1)^{2j}\chi_j(\tilde{q})\chi_{j0}(\tilde{q})\psi_0(\tilde{q}) =\\
&={k\over 2}\sum_{j,r}N^r_{a_1,a_2}\sum_{j',j'',n_1,n_2}{S_{rj}S_{jj'}S_{jj''}\over S_{0j}}(-1)^{2j}
\chi_{j'}(q){\chi_{j''n_1}(q)\psi_{n_2}(q)\over \sqrt{2k}\sqrt{2k}} \\
&={1\over 4}\sum_{j,r}N^r_{a_1,a_2}\sum_{j',j'',n_1,n_2}{S_{rj}S_{jj'}S_{j,{k\over 2}-j''}\over S_{0j}}
\chi_{j'}(q)\chi_{j''n_1}(q)\psi_{n_2}(q) \\
&={1\over 4}\sum_{r}\sum_{j',j'',n_1,n_2}N^r_{a_1,a_2}N^{j'}_{r,{k\over 2}-j''}\chi_{j'}(q)\chi_{j''n_1}(q)\psi_{n_2}(q) \\
&={1\over 4}\sum_{r}\sum_{j',j'',n_1,n_2}N^r_{a_1,a_2}N^{j'}_{r,{k\over 2}-j''}\chi_{j'}(q)\chi_{{k\over 2}-j'',k+n_1}(q)\psi_{n_2}(q) \\
&={1\over 4}\sum_{r}\sum_{j',j'',n_1,n_2}N^r_{a_1,a_2}N^{j'}_{r,j''}\chi_{j'}(q)\chi_{j'',n_1}(q)\psi_{n_2}(q)
\end{aligned}
\end{equation}
Here passing from the first to the second line we used the symmetry property (\ref{prsymod}) of the $SU(2)_k$
modular transformation matrix, and passing from the third to the forth line the field identification property
of the parafermion primaries $(j,n)\sim (k/2-j,k+n)$.  
We see that contribution of the first and the second terms are equal, and therefore 
in the partition function calculations
we can effectively
replace the first projector by one. Using the same arguments ( symmetries of $S$-matrix and the field identification)
we can show that also the second projector can be effectively replaced by one.
Taking this into account for the second part in the parenthesis we obtain:
\begin{equation}
\label{parsu3}
\begin{aligned}
k\sum_{j,r}N^r_{a_1,a_2}&{S_{rj}\over S_{0j}}\chi_j(\tilde{q})\chi_{jk}(\tilde{q})\psi_k(\tilde{q}) =\\
& =k\sum_{j,r}N^r_{a_1,a_2}\sum_{j',j'',n_1,n_2}{S_{rj}S_{jj'}S_{jj''}\over S_{0j}}
\chi_{j'}(q){\chi_{j''n_1}(q)\psi_{n_2}(q)\over \sqrt{2k}\sqrt{2k}}(-1)^{(n_1+n_2)} \\
&={1\over 2}\sum_{r}\sum_{j',j'',n_1,n_2}N^r_{a_1,a_2}N^{j'}_{r,j''}\chi_{j'}(q)\chi_{j''n_1}(q)\psi_{n_2}(q)(-1)^{(n_1+n_2)}
\end{aligned}
\end{equation}
Collecting (\ref{parsu}), (\ref{parsu2}) and (\ref{parsu3}) we obtain (\ref{pf1}).

For boundary states (\ref{carsttw}) and (\ref{seccar}) using (\ref{perdec}) we have
\begin{eqnarray}
&&{}_{C}^{(1)}\langle a_1|(\tilde{q})^{L_0-c/24}|a_2\rangle_{\cal P}=
\sqrt{k}\sum_{j,n}{S_{a_1j}S_{a_2j}\over S_{0j}S_{0j}}\chi_{j}(\tilde{q}){1+(-1)^{2j+n}\over 2}\times\nonumber \\
&&{}^{0\bar{1}}\langle\langle Bj|(\tilde{q})^{L_0-c/24}
|jn\rangle\rangle^{PF_0\times PF_{\bar{1}}}_{\tau}|n\rangle\rangle^{U(1)_0\times U(1)_{\bar{1}}}_{\tau+}
\end{eqnarray}
Let us recall that
\begin{eqnarray}
\label{parnd}
{}^{U(1)_0\times U(1)_{\bar{1}}}{}_{\tau-}\langle\langle n|(\tilde{q})^{L_0-c/24}|r\rangle\rangle^{U(1)_0\times U(1)_{\bar{1}}}_{\tau+}
=\delta_{n,0}\delta_{r,0}\chi_{ND}(\tilde{q})
\end{eqnarray}
where 
\begin{eqnarray}
\chi_{ND}(\tilde{q})=(\tilde{q})^{-1/24}{\rm Tr}(P(\tilde{q})^{L_0^{U(1)}})={1\over (\tilde{q})^{1/24}\prod(1+(\tilde{q})^n)}
\end{eqnarray}
where $P$ takes $X$ to $-X$.
This can be derived noting that zero-modes do not contribute, and the overlap is given
by the parity-weighted partition function receiving contribution only from higher oscillator states.
Using (\ref{parnd}) we obtain 
\begin{eqnarray}
{}_{C}^{(1)}\langle a_1|(\tilde{q})^{L_0-c/24}|a_2\rangle_{\cal P}=
\sqrt{k}\sum_{j}{S_{a_1j}S_{a_2j}\over S_{0j}S_{0j}}{1+(-1)^{2j}\over 2}\chi_{j}(\tilde{q})\chi_{j0}(\tilde{q})\chi_{ND}(\tilde{q})
\end{eqnarray}
Performing modular transformation and using that
\begin{eqnarray}
\chi_{ND}(\tilde{q})=\sqrt{2}{q^{1/48}\over \prod(1-q^{n-1/2})}
\end{eqnarray}
we obtain (\ref{pf2})
\begin{eqnarray}
&&{}_{C}^{(1)}\langle a_1|(\tilde{q})^{L_0-c/24}|a_2\rangle_{\cal P}=\sqrt{k}\sum_{r}\sum_{j',j'',n}
N^r_{a_1a_2}N^{j'}_{rj''}\chi_{j'}(q){\chi_{j''n}(q)\over\sqrt{2k}}\sqrt{2}{q^{1/48}\over \prod(1-q^{n-1/2})}=
\nonumber \\
&&\sum_{r}\sum_{j',j'',n}
N^r_{a_1a_2}N^{j'}_{rj''}\chi_{j'}(q)\chi_{j''n}(q){q^{1/48}\over \prod(1-q^{n-1/2})}
\end{eqnarray}

For boundary states (\ref{carst}) and (\ref{seccar}),
again using permuted decomposition (\ref{perdec}) and similar for $\bar{0}1$ sector
we can write
\begin{eqnarray}
\label{ovmmus}
&&{}^{(1)}_C\langle a|(\tilde{q})^{L_0-c/24}|a_0,a_1\rangle=
\sqrt{k}\sum_{j,j_0,j_1}\sum_{n_1,n_2,n_3=1}^{2k}\sum_{i=0,k}
{S_{aj}\over S_{0j}}{S_{a_0j_0}S_{a_1j_1}\over \sqrt{S_{0j_0}S_{0j_1}}}\times\nonumber\\
&&A^{j,j_0,j_1}_{n_1,n_2,n_3,i}B_{n_1,n_2,n_3,i}{1+(-1)^{2j_0+n_1}\over 2}{1+(-1)^{2j_1+n_2}\over 2}{1+(-1)^{2j+n_3}\over 2}
{1+(-1)^{2j+i}\over 2}
\end{eqnarray}
where $A^{j,j_0,j_1}_{n_1,n_2,n_3,i}$ and $B_{n_1,n_2,n_3,i}$  are contributions from the parafermion and $U(1)_k$ parts respectively
and given by the expressions
\begin{eqnarray}
\label{aaa}
A^{j,j_0,j_1}_{n_1,n_2,n_3,i}={}^{0\bar{1}}\langle\langle j,i|{}^{\bar{0}1}\langle\langle j,n_3|
(\tilde{q})^{L_0^{PF}-c/24}|j_0,n_1\rangle\rangle^{PF_0}_u|j_1,n_2\rangle\rangle^{PF_1}_u
\end{eqnarray}
\begin{eqnarray}
\label{uparov}
B_{n_1,n_2,n_3,i}={}^{0\bar{1}}{}_{\tau-}\langle\langle i|{}^{\bar{0}1}{}_{\tau+}\langle\langle n_3|
(\tilde{q})^{L_0^{U(1)}-c/24}|An_1\rangle\rangle^{U(1)_0}_u|An_2\rangle\rangle^{U(1)_1}_u
\end{eqnarray}

Following the same steps as in formula (\ref{ishot}) for (\ref{aaa}) we can write
\begin{eqnarray}
\label{apart}
A^{j,j_0,j_1}_{n_1,n_2,n_3,i}=\chi_{j,n}(\tilde{q}^2)\delta_{j,j_0}\delta_{j,j_1}\delta_{i,n_1}\delta_{i,n_2}\delta_{i,n_3}
\end{eqnarray}

Calculating (\ref{uparov}) we again can follow the similar steps as in the formula
(\ref{ishot}), but presence of ``$\tau-$'' state in
(\ref{uparov}) will bring the following changes.  For $U(1)_k$
boundary states as we can see from (\ref{untwishd}), (\ref{untwishn}),
(\ref{ishstg}) and (\ref{ishstgg}) the index labeling orthonormal
basis of the $U(1)_k$ representations runs over zero-mode momentum
part, and over $\alpha$'s created part.  Denoting the zero-mode
momenta in four states in (\ref{uparov}) by $p^{0\bar{1}}$,
$p^{\bar{0}1}$,$p^{0\bar{0}}$,$p^{1\bar{1}}$ respectively by the
similar steps as in (\ref{ishot}) we obtain:
\begin{equation}
\begin{aligned}
&p^{0\bar{1}}=p^{0\bar{0}}\, , \quad
-p^{0\bar{1}}=p^{1\bar{1}} \\
&p^{\bar{0}1}=p^{0\bar{0}}\, \quad 
p^{\bar{0}1}=p^{1\bar{1}} \, , 
\end{aligned}
\end{equation}
where the minus in the second line comes from the ``$\tau-$'' type first Ishibashi state in (\ref{uparov}).
It is easy to see that these four conditions imply that all zero-modes momenta are zero.
Therefore as before zero-modes do not contribute to  (\ref{uparov}).
For the $\alpha$'s created part we again get as in (\ref{ishot}) doubled energy in exponent
but weighted with the sign coming from  the
`$\tau-$'' type Ishibashi state:
\begin{eqnarray}
\label{uparovv}
B_{n_1,n_2,n_3,i}=(\tilde{q})^{-1/12}{\rm Tr}(P(\tilde{q})^{2L_0^{U(1)}})\delta_{n_1,0}\delta_{n_2,0}\delta_{n_3,0}
\delta_{i,0}=\chi_{ND}(\tilde{q}^2)\delta_{n_1,0}\delta_{n_2,0}\delta_{n_3,0}\delta_{i,0}
\end{eqnarray}
 Putting  (\ref{apart}) and (\ref{uparovv}) in (\ref{ovmmus}) we get:
\begin{eqnarray}
&&{}^{(1)}_C\langle a|(\tilde{q})^{L_0-c/24}|a_0,a_1\rangle=
\sqrt{k}\sum_{j,r}N^r_{a_0a_1}S_{rj}
{S_{aj}\over S_{0j}}{1+(-1)^{2j}\over 2}\chi_{j,0}(\tilde{q}^2)\chi_{ND}(\tilde{q}^2)=\nonumber\\
&&\sqrt{k}\sum_{r,j,j',n}N^r_{a_0a_1}{S_{rj}S_{aj}S_{jj'}\over S_{0j}}{1+(-1)^{2j}\over 2}
{\chi_{j',n}\over\sqrt{2k}}(q^{1/2})\sqrt{2}{(q^{1/2})^{1/48}\over \prod(1-(q^{1/2})^{n-1/2})}=\nonumber\\
&&\sum_{r,j',n}N^r_{a_0a_1}N^{j'}_{ra}
\chi_{j',n}(q^{1/2}){(q^{1/2})^{1/48}\over \prod(1-(q^{1/2})^{n-1/2})}
\end{eqnarray}
which is (\ref{pf3}).

For boundary states (\ref{thircar2}) we have
\begin{eqnarray}
&&{}^{(3)}_C\langle a_1|(\tilde{q})^{L_0-c/24}|a_2\rangle^{(3)}_C=
2k\sum_{j,n}{S_{a_1j}S_{a_2j}\over S_{0j}S_{0j}}{1+(-1)^{2j+n}\over 2}\times\nonumber\\
&&\chi_{j,n}(\tilde{q})\chi_{j,n}(\tilde{q})\psi_n(\tilde{q})\psi_n(\tilde{q})
\end{eqnarray}
Performing modular transformation we obtain (\ref{pfff1})
\begin{eqnarray}
 &&Z_{a_1a_2}=
2k\sum_{r,n,j,j',j''}\sum_{n_1,n_2,n_3,n_4}N^r_{a_1a_2}{S_{rj}S_{jj'} S_{jj''}\over S_{0j}}
{\chi_{j',n_1}(q)\chi_{j'',n_2}(q)\psi_{n_3}(q)\psi_{n_4}(q)\over\sqrt{2k}\sqrt{2k}\sqrt{2k}\sqrt{2k}}
e^{i{\pi\over k}n(n_1+n_2+n_3+n_4)}=\nonumber\\
&&\sum_{r,j',j''}\sum_{n_2,n_3,n_4}N^r_{a_1a_2}
N^{j'}_{rj''}\chi_{j',n_2+n_3+n_4}(q)\chi_{j'',n_2}(q)\psi_{n_3}(q)\psi_{n_4}(q)
\end{eqnarray}
where we used that
\begin{eqnarray}
\label{exproj}
\sum_n e^{i{\pi\over k}nn'}=2k\delta_{n',0}
\end{eqnarray}
Now consider partition function between states (\ref{carsttw}) and (\ref{thircar2}).
We note that parafermions in both sides appear in the permuted way, therefore
for them following to (\ref{ishover}) we obtain $\chi_{j,n}(\tilde{q})\chi_{j,n}(\tilde{q})$.
But we see that $U(1)_k$ states appear in permuted way in one side, and unpermuted way
in other side, therefore for them following to (\ref{ishot})
we obtain $\psi_n(\tilde{q}^2)$.
Taking all this into account we get:
\begin{eqnarray}
{}^{(3)}_C\langle a_1|(\tilde{q})^{L_0-c/24}|a_2\rangle_{\cal P}=
\sqrt{2k}\sum_{j,n}{S_{a_1j}S_{a_2j}\over S_{0j}S_{0j}}{1+(-1)^{2j+n}\over 2}
\chi_{j,n}(\tilde{q})\chi_{j,n}(\tilde{q})\psi_n(\tilde{q}^2)
\end{eqnarray}
Performing modular transformation we obtain (\ref{pfff2})
\begin{eqnarray}
 &&Z_{a_1a_2}=
\sqrt{2k}\sum_{r,n,j,j',j''}\sum_{n_1,n_2,n_3}N^r_{a_1a_2}{S_{rj}S_{jj'} S_{jj''}\over S_{0j}}
{\chi_{j',n_1}(q)\chi_{j'',n_2}(q)\psi_{n_3}(q^{1/2})\over\sqrt{2k}\sqrt{2k}\sqrt{2k}}
e^{i{\pi\over k}n(n_1+n_2+n_3)}=\nonumber\\
&&\sum_{r,j',j''}\sum_{n_2,n_3}N^r_{a_1a_2}
N^{j'}_{rj''}\chi_{j',n_2+n_3}(q)\chi_{j'',n_2}(q)\psi_{n_3}(q^{1/2})
\end{eqnarray}
For partition function between the states (\ref{carst}) and (\ref{thircar2}), the
situation is opposite. Now we have parafermions in the permuted way
in one side, and unpermuted way in other side, therefore leading to $\chi_{j,n}(\tilde{q}^2)$.
But we have $U(1)_k$ states in unpermuted way on the both sides, therefore leading to $\psi_n(\tilde{q})\psi_n(\tilde{q})$.
Collecting all we get
\begin{eqnarray}
\label{ovtrp}
&&{}^{(3)}_C\langle a|(\tilde{q})^{L_0-c/24}|a_0,a_1\rangle=
\sqrt{2k}\sum_{j}\sum_n
{S_{aj}\over S_{0j}}{S_{a_0j}S_{a_1j}\over S_{0j}}\times \nonumber\\
&&\chi_{j,n}(\tilde{q}^2)\psi_n(\tilde{q})\psi_n(\tilde{q})
{1+(-1)^{2j+n}\over 2}
\end{eqnarray}
Performing modular transformation we obtain (\ref{pfff3})
\begin{eqnarray}
&& Z_{a,a_0a_1}=\sqrt{2k}\sum_{j,r,n}\sum_{n_1,n_2,n_3}N^r_{a_0a_1}{S_{rj}
S_{aj}S_{jj'}\over S_{0j}}{1+(-1)^{2j+n}\over 2}
{\chi_{j',n_1}(q^{1/2})\psi_{n_2}(q)\psi_{n_3}(q)\over \sqrt{2k}\sqrt{2k}\sqrt{2k}}\times \nonumber\\
&&e^{i{\pi\over k}n(n_1+n_2+n_3)}=\sum_{r,j'}\sum_{n_2,n_3}N^r_{a_0a_1}N^{j'}_{ra}\chi_{j',n_2+n_3}(q^{1/2})\psi_{n_2}(q)\psi_{n_3}(q)
\end{eqnarray}
In the case of partition function between (\ref{seccar}) and (\ref{thircar2})
we have parafermion in the permuted way on both sides, but $U(1)_k$ states in permuted way
on one side, and unpermuted way in other side, and additionly we should take
into account that in the $0\bar{1}$ sector of (\ref{seccar}) we have ``$\tau-$'' state, whereas
in (\ref{thircar2}) we have $A$ type states.
Therefore repeating the same steps as in  derivation of (\ref{uparovv})
we get
\begin{eqnarray}
 Z_{a_1a_2}=
\sqrt{2k}\sqrt{k}\sum_{j}{S_{a_1j}S_{a_2j}\over S_{0j}S_{0j}}{1+(-1)^{2j}\over 2}
\chi_{j,0}(\tilde{q})\chi_{j,0}(\tilde{q})\chi_{ND}(\tilde{q}^2)
\end{eqnarray}
Performing modular transformation we get (\ref{pfff4})
\begin{eqnarray}
&& Z_{a_1a_2}=\sqrt{2k}\sqrt{k}\sum_{j,r,j',j''}\sum_{n_1,n_2}N^r_{a_1a_2}{S_{rj}
S_{j'j}S_{jj''}\over S_{0j}}{\chi_{j',n_1}(q)\chi_{j'',n_2}(q)\over \sqrt{2k}\sqrt{2k}}\sqrt{2}
{(q^{1/2})^{1/48}\over \prod(1-(q^{1/2})^{n-1/2})}=\nonumber\\
&&\sum_{j',j'',r}\sum_{n_1,n_2}N^r_{a_1a_2}N^{j'}_{rj''}\chi_{j',n_1}(q)\chi_{j'',n_2}(q)
{(q^{1/2})^{1/48}\over \prod(1-(q^{1/2})^{n-1/2})}
\end{eqnarray}
\section{Some facts about $U(1)_k$, $SU(2)_k$ and parafermion theories}
\label{paraf}
In this section we briefly review some necessary facts about 
$U(1)_k$, $SU(2)_k$ and ${\cal A}^{ PF(k)}={SU(2)_k\over U(1)_k}$ theories.
\medskip

\noindent {\bf $U(1)_k$ theory:}
\vskip5pt
\noindent The $U(1)_k$ chiral algebra ($k\in Z$) contains, besides the
Gaussian $U(1)$ current $J=i\sqrt{2k}\partial X$, two additional
generators
\begin{equation}
\label{adsym}
\Gamma^{\pm}=e^{\pm i\sqrt{2k}X}
\end{equation}
of integer dimension $k$ and charge $\pm 2k$.  The primary fields of
the extended theory are those vertex operators~$e^{i\gamma X}$ whose
OPEs with the generators~(\ref{adsym}) are local.  This fixes~$\gamma$
to be
\begin{equation}
\label{verop}
\gamma={n\over \sqrt{2k}}\,,\quad n\in Z
\end{equation}
Their conformal dimension is $\Delta_n={n^2\over 4k}$.  For primary
fields, the range of~$n$ must be restricted to the fundamental domain
\mbox{$n=-k+1,-k+2,\ldots,k$} since a shift of~$n$ by~$2k$ in~$e^{in
X/\sqrt{2k}}$ amounts to an insertion of the ladder
operator~$\Gamma^{+}$, which thereby produces a descendant field.

From the point of view of the extended algebra the characters are
easily derived. A factor $q^{\Delta_n-1/24}/\eta(q)$ takes care of the
action of the free boson generators. To account for the effect of the
distinct multiple applications of the generators~(\ref{adsym}), which
yield shifts of the momentum~$n$ by integer multiples of~$2k$, we must
replace~$n$ by~$n+l2k$ and sum over~$l$. The net result is
\begin{equation}
\label{char}
\psi_n(q)={1\over \eta(q)}\sum_{l\in Z} q^{k(l+n/2k)^2} \, .
\end{equation}
The action of the modular transformation $S$ on the characters (\ref{char}) is
\begin{equation}
\label{modtr}
\psi_n(q')={1\over \sqrt{2k}}\sum_{n'}e^{{-i\pi nn'\over
k}}\psi_{n'}(q) \quad q=e^{2\pi i\tau} \quad \tau'=-{1\over \tau} \, .
\end{equation}
\medskip

\noindent {\bf The parafermion ${\cal A}^{ PF(k)}={SU(2)_k\over U(1)_k}$} 
\vskip10pt 
\noindent The chiral algebra of this theory has a set of irreducible
representations described by pairs $(j,n)$ where $j\in {1\over 2}Z$,
$0\leq j \leq k/2$, and $n$ is an integer defined modulo $2k$.  The
pairs are subject to a constraint~$2j+n=0 \mod 2$, and an
equivalence relation $(j,n)\sim (k/2-j,k+n)$. The character of the
representation $(j,n)$, denoted by $\chi_{j,n}(q)$, is determined
implicitly by the decomposition
\begin{equation}
\label{chardec}
\chi^{SU(2)}_j(q)=\sum_{n=-k}^{k+1}\chi_{j,n}^k(q)\psi_n(q) \, .
\end{equation}
The action of modular group on the character is
\begin{equation}
\chi_{j,n}^k(q')=\sum_{(j',n')}S^{PF}_{(j,n),(j'n')}\chi_{j',n'}^k(q)
\end{equation}
and the PF S-matrix is
\begin{equation}
\label{parmod}
S^{PF}_{(j,n),(j'n')}={1\over \sqrt{2k}}e^{{i\pi nn'\over k}}S_{jj'}\,,
\end{equation}
where $S_{jj'}$ defined in~(\ref{smatr}).

When combining left and right-movers, the simplest modular invariant
partition function of the parafermion theory is obtained by summing
over all distinct representations
\begin{equation}
\label{parpar}
Z=\sum_{(j,n)\in PF_k}|\chi_{j,n}|^2 \, .
\end{equation}
The parafermion theory has a global $Z_k$ symmetry under which the fields $\psi_{j,n}$
generating the representation $(j,n)$ transform as
\begin{equation}
\label{zksym}
g: \quad  \psi_{j,n}\rightarrow \omega^{n}\psi_{j,n}, \quad
\omega=e^{{2\pi i\over k}} \, .
\end{equation}
Therefore we can orbifold the theory by this group. Taking the symmetric
orbifold by $Z_k$ of~(\ref{parpar}) leads to the partition function
\begin{equation}
\label{parorb}
Z=\sum_{(j,n)\in PF_k}\chi_{j,n}\bar{\chi}_{j,-n} \, .
\end{equation}
We see that effect of the orbifold is to change the relative sign
between the left and right movers of the~$U(1)$ group with which we
orbifold. 
Therefore the $Z_k$ orbifold of the parafermion theory at level~$k$ is
T-dual to the original theory. This fact will be the basis of many
constructions in the main text.

\section{Various coordinate systems for the sphere and relations between them}

A three-sphere $S^3$ is a group manifold of the $SU(2)$ group. A
generic element in this group can be written as
\begin{equation}
\label{SU(2)}
g =  \, X_0 \sigma_0 + i (X_1 \sigma_1 + X_2 \sigma_2 + X_3
\sigma_3) = \begin{pmatrix}
X_0 + i X_3 \, & \,  X_2 + iX_1 \\
-(X_2 - i X_1)  \, &  \, X_0 - iX_3 
\end{pmatrix}
\end{equation}
subject to condition that the determinant  is equal to one
\begin{equation}
\label{hyper}
X_{0}^2 + X_1^2 + X_2^2+X_{3}^2 = 1 \, .
\end{equation}
The metric on $S^3$ can be written in the following three ways, which
will be used in the main text. Firstly, using the Euler parametrisation
of the group element we have
\begin{eqnarray}
\label{euler-s3}
g &=& e^{i\chi {\sigma_3 \over 2}}e^{i \tilde{\theta} {\sigma_1 \over 2}} e^{i \varphi {\sigma_3 \over 2}} \, \\ 
{\rm d}s^2 &=& {1\over 4} \Big( ({\rm d} \chi + \cos \tilde{\theta} {\rm d} \varphi)^2 + {\rm d} \tilde{\theta}^2 + \sin^2 \tilde{\theta} {\rm d} \varphi^2 \Big) \, .
\end{eqnarray}
The ranges of coordinates are $0 \leq \tilde{\theta}\, \leq \pi$,  $0 \leq \varphi \leq 2 \pi$ and $0 \leq \chi \leq 4 \pi$.

Secondly, we can use coordinates that are analogue to the global
coordinate for~$AdS_3$
\begin{eqnarray}
\label{global-s3}
X_0+iX_3 &=& \cos \theta e^{i\tilde{\phi}} \, , \quad X_2+iX_1 = \sin \theta e^{i\phi}\, \\ 
{\rm d}s^2 &=& {\rm d} \theta^2 + \cos^2 \theta {\rm d} \tilde{\phi}^2 + \sin^2 \theta {\rm d} \phi^2 \, .  
\end{eqnarray}
The relation between the metrics~(\ref{euler-s3}) and~(\ref{global-s3}) is given by
\begin{equation}
\label{glob-par}
\chi = \tilde{\phi} + \phi  \, , \quad \varphi = \tilde{\phi} - \phi \, ,
 \quad \theta={\tilde{\theta} \over 2} \,.
\end{equation}
The ranges of coordinates are $-\pi\leq\tilde{\phi},\phi\leq\pi$ and
$0\leq\theta\leq{\pi\over 2}$.

Thirdly, the standard metric on~$S^3$ is given by ($\vec{n}$ is a unit
vector on~$S^2$)
\begin{eqnarray}
\label{standard-s3}
g &=& e^{2i  \psi {\vec{n} \cdot \vec{\sigma} \over 2}} \, , \quad
{\rm d}s^2 = {\rm d} \psi^2 + \sin^2 \psi ( {\rm d} \xi^2 + \sin^2 \xi {\rm d} \eta^2) \, \\
X_0+iX_3&=& \cos \psi +i\sin \psi\cos \xi \, , \quad X_2+iX_1 = \sin \psi \sin \xi e^{i\eta} \,.
\end{eqnarray}
The ranges of the coordinates are $0 \leq\psi\, ,\, \xi  \leq \pi$
and $ 0\leq \eta \leq  2 \pi$.

\bibliographystyle{JHEP} 
\bibliography{grpr}
\end{document}